\documentclass[oneside]{article}

\usepackage{amsmath,amsfonts }
\usepackage{color}
\usepackage{autobreak}
\usepackage{graphicx}
\usepackage{hyperref}
\usepackage{amssymb}
\usepackage{multicol}
\usepackage{multirow}
\usepackage{color, colortbl}
\usepackage{makecell, caption, booktabs, multirow, siunitx}
\usepackage{booktabs}

\usepackage{float}
\usepackage{setspace}

\usepackage{inputenc}

\definecolor{Gray}{gray}{0.9}

\newcommand {\db} {\boldsymbol{d}}
\newcommand {\eb} {\boldsymbol{e}}
\newcommand {\rb} {\boldsymbol{r}}

\newcommand {\Fb} {\boldsymbol{F}}

\newcommand {\Db} {\boldsymbol{D}}

\newcommand {\K} {\mathcal{\kappa}}

\newcommand{\uu} {\boldsymbol{\mathsf{u}}}
\newcommand{\vv} {\boldsymbol{\mathsf{v}}}
\newcommand{\ww} {\boldsymbol{\mathsf{w}}}
\newcommand{\oo} {\boldsymbol{\mathsf{o}}}
\newcommand{\Ec} {\mathcal{E}}
\newcommand{\NN} {\mathcal{N}}
\newcommand{\CC} {\mathsf{C}}
\newcommand{\DD} {\Gamma}
\newcommand{\BB} {{\Lambda}}
\usepackage{todonotes}
\usepackage{calrsfs}

\begin{document}
	
\doublespacing

\title{\vspace{-3cm} {\bf  Competition between epithelial tissue elasticity and surface tension\\ in cancer  morphogenesis}}

\author{
 Antonino Favata$^1$ \and Roberto Paroni$^{2}$ \and Filippo Recrosi$^3$ \and Giuseppe Tomassetti$^4$
}

%

\maketitle

\vspace{-1cm}
\begin{center}
	{\small
	
		$^1$ Department of Structural and Geotechnical Engineering\\
		Sapienza University of Rome, Rome, Italy\\
		\href{mailto:antonino.favata@uniroma1.it}{antonino.favata@uniroma1.it}\\[8pt]

		$^2$ Dipartimento di Ingegneria Civile e Industrialeg\\
		Universit\`{a} di Pisa, Pisa, Italy\\
		\href{mailto:roberto.paroni@unipi.it}{roberto.paroni@unipi.it}\\[8pt]

	$^3$ Department of Structural and Geotechnical Engineering\\
	Sapienza University of Rome, Rome, Italy\\
	\href{mailto:filippo.recrosi@uniroma1.it}{filippo.recrosi@uniroma1.it}\\[8pt]
	
	$^4$ Department of Engineering\\
	 Roma Tre University\\
	\href{mailto:giuseppe.tomassetti@uniroma3.it}{giuseppe.tomassetti@uniroma3.it}\\[8pt]

	}
\end{center}

\pagestyle{myheadings}
\markboth{A.~Favata, R.~Paroni, F.~Recrosi, G.~Tomassetti}
{Epithelial tissue elasticity and surface tension}

\vspace{-0.5cm}
\section*{Abstract}
We derive a continuum mechanical model to capture the morphological changes occurring at the pretumoral stage of  epithelial tissues. The proposed model aims to investigate the competition between the bulk elasticity of the epithelium and the surface tensions of the apical and basal sides. According to this model, when the apico-basal tension imbalance reaches a critical value, a subcritical bifurcation is triggered and the epithelium attains its physiological folded shape. Based on data available in the literature, our model predicts that pretumoral cells are softer than healthy cells. 

\vspace{1cm}

\noindent {\bf Keywords}: Epithelium, subcritical bifurcation, surface energy, cancer morphogenesis, rod theories, cell elasticity

\tableofcontents

\vspace{.5cm}

        \section{Introduction}

\subsection{Physiology of epithelial tissues}

Epithelial tissues are one of the most widespread type of tissue in living things. Epithelia differ in shape and function. They appear in mono or multi-layers of cells covering and protecting the inner parts of tissues and organs. These layers are in the shape of flat sheets in the case of skin, or in the shape of corrugated and folded membranes in stomach and intestine, where they give rise to villi and crypts. These corrugations increase the exchanging area, favoring the secretion of enzymes and absorption of nutrients. During embryogenesis epithelial tissues differentiate from all the three embryonic cell layers, undergoing extensive and precise morphological changes, which result in complex folding patterns.

Epithelial morphogenesis is characterized by a highly complex chemo-mechanical phenomenology which is not comprehensively understood yet, and whose review is out of topic of the present paper. Despite this complexity,  the purely mechanical aspects of these processes are key in determining tissural architecture. In particular, it is known that epithelial folding can be the manifestation of a mechanical instability triggered by the contractile action of a meshwork of cross-linked actin filaments acting in the proximity of the apical and basal membranes of the cells (see the sketch in  Fig. \ref{fig:meshwork} from Section \ref{sec:energetics} of the present paper).

Epithelia are a common site of tumor onset: Carcinomas, arising from the epithelium, represent  more than 80\% of the cancer-related deaths in the Western world  \cite{Weinberg-2013}. In particular, Pancreatic Ductal Adenocarcinoma (PDAC) is the most lethal of the common cancers, without effective therapeutic option except surgery \cite{Therville-2019}.


\subsection{Brief survey of available mechanical models}
Perhaps the earliest attempt to understand the mechanical basis of epithelial folding dates back to the work of W.H. Lewis \cite{lewis_mechanics_1947}, who came up with a two-dimensional physical model consisting of a linear framework of pinned bars, representing an epithelium sheet, with elastic cables stretched on the top and bottom side, which would mimic the apical and the basal tensions. More recently, several mechanical models have been devised: from the 2D and 3D vertex models \cite{Honda-1980, Nagai-Honda-2009, Bergmann-2018,latorre_active_2018,misra_shape_2016,misra_complex_2017}, to continuum models \cite{Murisic-2015, Bielmeier-2016}, which picture an epithelial layer as a planar rod or as a shell. Among of the latter, some recent works investigate the epithelial tissue morphogenesis as triggered by buckling instability \cite{Drasdo-2000, Shraiman-2005, Hohlfeld-2011, Li-2012,BenAmar-2013,Balbi-2015, Salbreux-2017,Destrade_2020} or by differential intraepithelial tensions \cite{Hannezo-2014, Krajnc-2013,Papastavrou-2013,Julicher-2018}. 
In \cite{Krajnc-2013, Krajnc-2015, Krajnc-2016} apico-basal differential tension is shown to be enough to produce folded configurations in longitudinal epithelial sheets. In \cite{Krajnc-2015} a continuum model, derived from area- and perimeter-elasticity (APE) models \cite{Farhadifar-2007}, is proposed for the healthy epithelium.

Concerning the connection with tumor morphogenesis, as reviewed in \cite{Krajnc-2020}, the available models address multiple mechanical factors as responsible for the disruption of normal tissue architecture besides the abnormal acto-myosin concentration gradient from basal to apical region \cite{Messal}, which is the mechanical effect that we address in the present paper; in particular, other factors include cancer cells proliferation \cite{Bielmeier-2016}, lateral cell adhesion \cite{Hannezo-2014}, elasticity of the basement membrane \cite{Krajnc-2015}.

\subsection{The proposed model and its implications }
We model an epithelial monolayer as a two dimensional thin body  equipped with a bulk and a surface energy at the apical and basal sides.  These energetic terms are in competition: the former favors the undeformed configuration; the latter induces bending when the apical and basal energies are imbalanced. By dimension reduction, based on a kinematic Ansatz allowing for thickness extension, we arrive at a one-dimensional model of a nonlinear elastic rod whose equilibria are governed by the competition of the aforementioned energetic contributions.

In our model two dimensionless key parameters are introduced, $\gamma$ and $\sigma$: the former is a measure of the relative importance of surface energy compared to bulk energy; the latter is a measure of the imbalance between apical and basal tensions.

As $\gamma$ grows, surface energy becomes more important: the apical and basal sides shorten and, in turn, the thickness increases. A growth of $\sigma$ favors curved configurations.

We formulate a nonlinear equilibrium problem that admits in principle a manifold of solutions. The rectilinear configuration is a solution of this problem for every choice of $\gamma$ and $\sigma$. For $\gamma$ small, bulk elasticity prevails over surface tension and there is no other solution except the rectilinear one; for $\gamma$ large enough, there exists a critical value $\sigma_c$ of the parameter $\sigma$ where bifurcation from the rectilinear configuration occurs.

A careful analysis, performed through the Lyapunov-Schmidt decomposition, reveals that the bifurcation is subcritical. This is confirmed by our numerical calculations.



Our model predicts a distinctive mechanical behavior of pre-cancerous cells. Based on data available in \cite{Messal}, we estimate the pretumoral tissue softening for pancreatic Neoplasia. This result is in accordance with elastographic measures in \cite{Therville-2019}, and confirms that transformed cells are softer than healthy cells.

\subsection{Organization of the manuscript}
In Section 2 we derive our model by prescribing the geometry and the underlying kinematical hypotheses. We specify the form of the bulk and surface energy, and we introduce an incompressibility constraint. We identify the relevant dimensionless parameters, we derive the equilibrium equations, and we formulate a boundary-value problem.

In Section 3 we study the loss of positivity of the elasticity tensor  and we determine the critical value of the apico-basal tension imbalance, along with the bifurcation mode.

In Section 4 we perform a detailed bifurcation analysis by means of the Lyapunov-Schmidt decomposition, and we identify the type of bifurcation. We accompany our analysis with a numerical calculation.

In Section 5 we discuss the implication of our model concerning the softening which accompanies incipient tumorigenesis.

\section{Modeling}

\subsection{Geometry and kinematics}
In line with recent work addressing folding patterns through continuum models \cite{Krajnc-2013,haas2019}, we restrict our attention to planar deformations. In accordance with this point of view, we identify a single layer of epithelial cells with a thin strip $\Omega$ of length $\ell$ and finite thickness $h$, and we choose an Ansatz on the class of possible deformations which will leads us to a one-dimensional model of a rod deforming on a plane.

One of the features of our approach is that our Ansatz involves a scalar parameter $\mu$ (see \eqref{eq:6} below) which describes transverse extension/contraction of the epithelial sheet, as observed in experiments (see, \textit{e.g.}, \cite{Messal}). We take into account the overall incompressibility of the epithelial sheet  by enforcing volume conservation \emph{on average} along the thickness (see \eqref{eq:4} and \eqref{lm1} below). The resulting model predicts that the ability of the epithelium to undergo transversal stretching and contraction is a key ingredient for the bifurcations that mark the transition from flat to folded configurations. This idea is not new in mechanics: bending instabilities of rods accompanied by non-uniform striction and/or dilation of the transversal fibers manifest themselves with the \emph{Brazier effect} \cite{antman,antman-2005,ppg1982,coman_2017}.

\begin{figure}[h]
	\centering
	\includegraphics[scale=1]{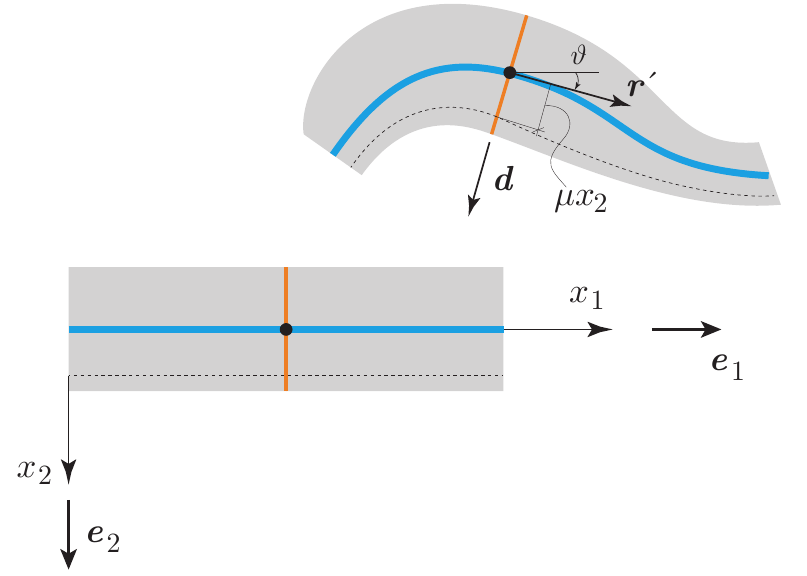}
	\caption{The strip in the reference and in the deformed configuration. As in standard rod theories, the transversal fiber (orange segment) remains straight and orthogonal to the mid axis (blue curve), but may undergo stretching.}
	\label{fig:trave}
\end{figure}

To describe our Ansatz, we introduce a coordinate system $(x_1,x_2)$, and we let $\{\eb_1,\eb_2\}$ be the associated orthonormal basis (see Fig.\ref{fig:trave}).
We assume that the deformation has the form:
\begin{equation}
{\boldsymbol f}\left(x_{1}, x_{2}\right)={\boldsymbol r}\left(x_{1}\right)+x_{2}{\boldsymbol d}\left(x_{1}\right).
\end{equation}
The vectors $\rb(x_1)$ and $\db(x_1)$ represent, respectively, the position of the midline of the epithelium and the orientation of the typical transversal fiber.
We allow the midline and the transverse direction to change length, thus the vectors $\rb'(x_1)$ and $\db(x_1)$  are not necessarily of unit length, but we assume that the transverse sections remain orthogonal to the midline curve:
$$
\rb'\cdot\db=0.
$$
Accordingly, the deformation gradient is 
\begin{equation}\label{F}
{\boldsymbol F}=
\nabla{\boldsymbol f}=\left({\boldsymbol r}^{\prime}+x_{2} {\boldsymbol d}^{\prime}\right) \otimes {\boldsymbol e_{1}}+{\boldsymbol d} \otimes {\boldsymbol e}_{2}.
\end{equation}

We denote by $\vartheta(x_1)$ the  rotation of the director $\db(x_1)$ with respect to the reference configuration. Then the unit vectors
\begin{equation}\label{a}
{\boldsymbol a}_{1}:=\frac{{\boldsymbol r}^{\prime}}{\left|{\boldsymbol r}^{\prime}\right|}=\cos \vartheta {\boldsymbol e}_{1}+\operatorname{sin} \vartheta {\boldsymbol e}_{2},\qquad \text{and}\qquad {\boldsymbol a}_{2}:=-\operatorname{sin} \vartheta {\boldsymbol e}_{1}+\cos \vartheta {\boldsymbol e}_{2}.
\end{equation}
represent, respectively, the tangent and the normal to the axis in the deformed configuration; a prime denotes differentiation with respect to the coordinate $x_1$. Thus, we can write
\begin{equation}\label{eq:6}
{\boldsymbol r}^{\prime}=\lambda {\boldsymbol a}_{1},\qquad {\boldsymbol d}=\mu{\boldsymbol a}_{2},
\end{equation}
where $\lambda$ and $\mu$ are, respectively, the \textit{axial} and \textit{transverse stretch}.     

From \eqref{a}, ${\boldsymbol a}_{2}'=-\vartheta^{\prime}{\boldsymbol a}_{1}$, hence $\db'=\mu'{\boldsymbol a}_{2}-\mu \vartheta^{\prime}{\boldsymbol a}_{1}$ and the deformation gradient given in \eqref{F} can be rewritten as
\begin{equation}\label{F2}
{\boldsymbol F}=\left(\lambda-x_{2} \mu \vartheta^{\prime}\right) {\boldsymbol a}_{1}\otimes {\boldsymbol e}_{1}+ x_{2} \mu^{\prime} {\boldsymbol a}_{2} \otimes {\boldsymbol e}_{1}+\mu {\boldsymbol a}_{2} \otimes {\boldsymbol e}_{2}.
\end{equation}

We next introduce the local rotation:
\begin{equation}
{\boldsymbol R}:={\boldsymbol a}_{1} \otimes {\boldsymbol e}_{1}+{\boldsymbol a}_{2} \otimes {\boldsymbol e}_{2}
\end{equation}
which maps the reference basis $\{\eb_1,\eb_2\}$ onto the current basis $\{{\boldsymbol a}_1,{\boldsymbol a}_2\}$.
Since we have in mind to deduce a beam-like model, we choose as deformation measure
\begin{equation}
{\boldsymbol D}:={\boldsymbol R}^{\top} {\boldsymbol F}=\left(\lambda-x_{2} \mu\vartheta^{\prime}\right){\boldsymbol e}_1 \otimes {\boldsymbol e}_{1}+x_{2} \mu^{\prime} {\boldsymbol e}_{2} \otimes {\boldsymbol e}_{1}+\mu {\boldsymbol e}_{2} \otimes {\boldsymbol e}_{2}.
\end{equation}
The strain measure $\Db$ makes free the deformation gradient $\Fb$ from the rigid rotation of the axis.
In the above equation, the last term on the right-hand side is the \textit{transverse stretch}. The first term is the sum of an \textit{average axial stretch} $\lambda {\boldsymbol e}_1\otimes {\eb_1}$, plus a linear term proportional to the \textit{curvature} $\vartheta'$ of the axis, weighted by $x_2\mu$, the latter representing the distance from the axis in the \emph{deformed configuration} (see Fig. \ref{fig:trave}). The second term describes a \textit{non-uniform shear} deformation associated to a possible non-uniformity of the transverse stretch.


\subsection{Energetics} \label{sec:energetics}
In our model we incorporate two different energetic contributions: a \textit{bulk energy} and a \textit{surface energy}. The former accounts for mechanical response of the cytoplasm, here assumed to be elastic; the latter takes into account the \emph{contractile tension} due to the thin meshwork of actin filaments laying beneath the cell membrane, depicted in the cartoons (b) and (c) in Fig. \ref{fig:meshwork} below.
\begin{figure}[h]\label{fig:meshwork}
	\centering
	\includegraphics[width=0.9\linewidth]{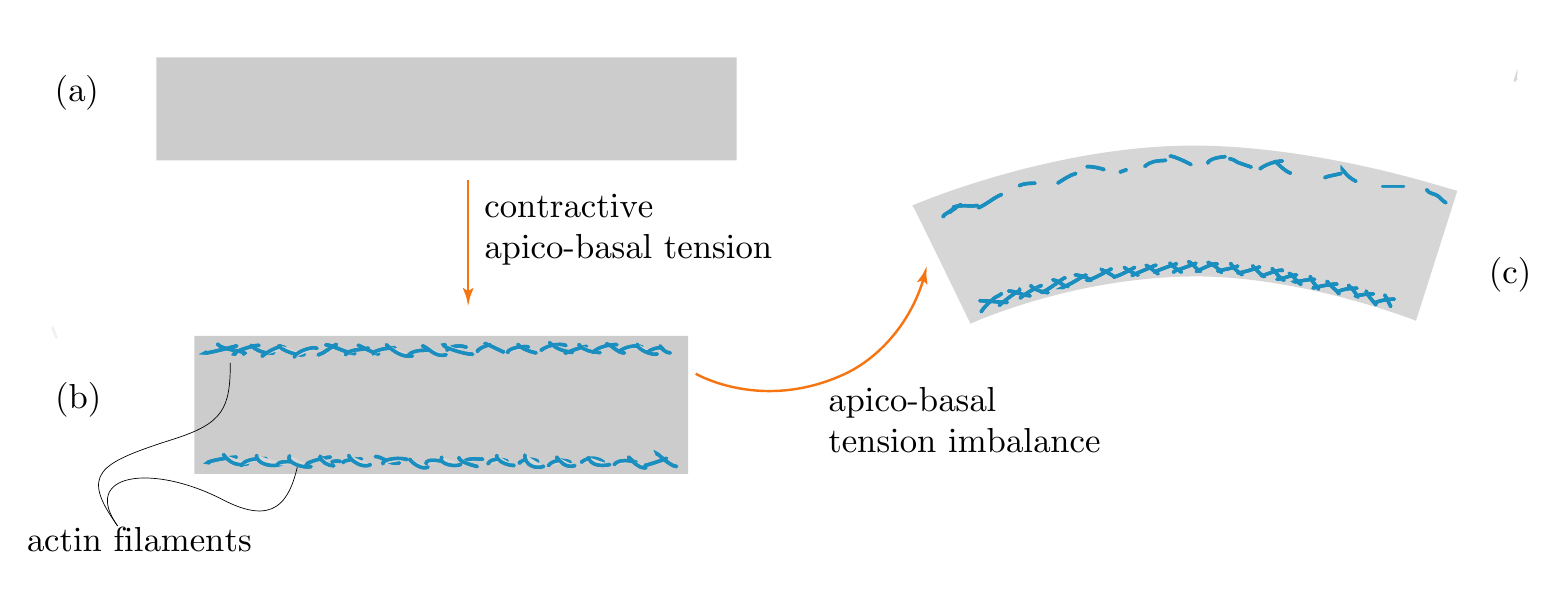}
	\caption{In the absence of surface tension and external load the strip is in equilibrium in the reference configuration (a). The balanced contractile tension of the actin filaments on the upper and lower side the strip induces lateral contraction, and in turn  transverse stretching due to incompressibility (b). Imbalance between contractile tension on the upper and lower sides of the strip results in the bent shape (c).}
	\label{fig:meshwork}
\end{figure}
In line with existing discrete models which consider an epithelial monolayer as a polygonal tessellation where each side of a polygon carries an energy proportional to its length, we assume that the surface energy be proportional to the tangential stretch at the boundary. This point of view has been already applied in \cite{Krajnc-2013} and \cite{haas2019} to derive continuum models for epitelial monolayers.

The consequences of including surface tension on the boundary of an elastic body have already been explored \cite{mora2013,bico2018,Papastavrou-2013}, and it is known that it can generate relevant mechanical effects on soft elastic bodies at small scales. In the present setting, however, we allow for an imbalance between the internal forces localized on the apical (top) and basal (bottom) sides of the strip which produces bending, as schematically depicted in Fig. \ref{fig:meshwork}-c.


We assume that the material is homogeneous and isotropic, and that it is in its natural state in the reference configuration. 
As a result, approximating the bulk energy with its Taylor expansion up to second order, we obtain:
\begin{equation}
W_b({\boldsymbol F})=W_b({\boldsymbol D})\simeq \frac{1}{2} D^{2} W_b({\boldsymbol I})[\text {sym }({\boldsymbol D}-\boldsymbol I), \text { sym } ({\boldsymbol D}-\boldsymbol I)];
\end{equation}
the terms $W_b({\boldsymbol I})$ and $DW_b({\boldsymbol I})$ do not appear since the reference is assumed to be natural, while the dependence on the symmetric part of $ {\boldsymbol D}-\boldsymbol I$ follows from the frame indifference of the energy density $W_b$.

The assumption that the bulk energy is isotropic entails that there exist constants $\alpha_1$ and $\alpha_2$ such that
\begin{equation}
\frac{1}{2} D^{2} W_b({\boldsymbol I})[\text {sym }({\boldsymbol D}-\boldsymbol I), \text { sym } ({\boldsymbol D}-\boldsymbol I)]=\alpha_{1}|\text { sym } ({\boldsymbol D}-\boldsymbol I)|^{2}+\alpha_{2}({\rm tr}  ({\boldsymbol D}-\boldsymbol I))^{2}.
\end{equation}
Hereafter, we assume the two material constants $\alpha_1$ and $\alpha_2$ to be strictly positive. 
The bulk strain energy per unit length along the direction $x_1$ is 
\begin{align}
\begin{autobreak}
\int_{-h/2}^{+h/2}W_b({\boldsymbol D}){\rm d}x_2= {w}_b,
\end{autobreak}
\end{align}
where
\begin{multline}
{w}_b(\lambda,\mu,\vartheta')= h\left[\alpha_{1}\left(\lambda^{2}+\mu^{2}\right)+\alpha_{2}(\lambda+\mu)^{2}-(\alpha_1+4\alpha_2)(\lambda+\mu)\right]\\
\qquad +\frac{h^{3}}{12}\left[\alpha_{1}\left(\mu^{2} \vartheta^{\prime 2}+\frac {{\mu'}^{2}}2\right)+\alpha_{2} \mu^{2} {\vartheta^\prime}^{2}\right].
\end{multline}
The first term in the above equation, proportional to the thickness, is a \emph{stretching energy}; the second term, which scales as the third power of the thickness, is a \emph{bending energy}.

In order to introduce the surface energy, let us consider an infinitesimal fiber parallel to the ${\boldsymbol e}_1$ direction in the reference configuration. The deformation transforms this fiber into, see \eqref{F2},
\begin{equation}
{\boldsymbol F}{\boldsymbol e}_1=\left(\lambda-x_{2} \mu \vartheta^{\prime}\right) {\boldsymbol a}_{1}+x_{2} \mu^{\prime} {\boldsymbol a}_{2}.
\end{equation}

Accordingly, the length of the generic fiber in the deformed configuration is
\begin{equation}\label{abstretch}
\left|{\boldsymbol F}{\boldsymbol e}_1\right|=\sqrt{\left(\lambda-x_{2} \mu\vartheta^{\prime}\right)^{2}+\left(x_{2} \mu^{\prime}\right)^{2}}.
\end{equation}

The surface energy is introduced as the cost paid to stretch the apical and basal fibers, \textit{i.e.}, $x_2=\mp h/2$, respectively. On denoting by  $\sigma_a$ and $\sigma_b$  the apical and basal tensions, \eqref{abstretch} allows to define the surface energy per unit length as follows:
\begin{equation}
{w}_{s}(\lambda,\mu,\vartheta')=\sigma_{a} \sqrt{\left(\lambda+\frac{h}{2} \mu \vartheta^{\prime}\right)^{2}+\frac{h^{2} \mu^{\prime^{2}}}{4}}+\sigma_{b} \sqrt{\left(\lambda-\frac{h}{2} \mu \vartheta^{\prime}\right)^{2}+\frac{h^{2} \mu^{\prime 2}}{4}}.
\end{equation}
We assume $\sigma_a$ and $\sigma_b$ to be non-negative.

\subsection{Incompressibility}
The cells in the epithelial monolayer, being filled with fluid, should be treated as  incompressible \cite{Krajnc-2013,haas2019}. Since each cell spans the entire thickness of the body, we find it appropriate to impose incompressibility on average over the thickness:
\begin{equation}\label{eq:4}
\frac 1 h\int_{-h/2}^{+h/2}\det \Fb{\rm d}x_2=1,
\end{equation}
where
\begin{equation}\label{eq:3}
\text { det } {\boldsymbol F}={\boldsymbol F}{{\boldsymbol e}_{1}} \times {\boldsymbol F}{\boldsymbol e}_2 \cdot {\boldsymbol e}_{3}=\left({\boldsymbol r}^{\prime}+x_{2} {\boldsymbol d}^{\prime}\right) \times{\boldsymbol d} \cdot {\boldsymbol e}_{3}.
\end{equation}
By combining \eqref{eq:3} and \eqref{eq:4} we obtain that the product of the longitudinal and transversal stretching is equal to one:
\begin{equation}\label{lm1}
\lambda\mu=1.
\end{equation}
From this identity we find $\lambda=1/\mu$ and hence, by dropping additive constants, we can rewrite the energy densities in terms of $\mu$ and $\vartheta$, only:
\begin{equation}
\begin{aligned}
{w}_b(\mu,\mu',\vartheta')&=  h\left(\left(\alpha_{1}+\alpha_{2}\right)\left(\frac{1}{\mu^{2}}+\mu^{2}\right)-(\alpha_1+4\alpha_2)\left(\frac 1 \mu+\mu\right)\right)\\
&+\frac{h^{3}}{12}\left[(\alpha_{1}+\alpha_2)\mu^2\vartheta'^2+\alpha_1\frac{\mu^{\prime 2}}2\right]
\end{aligned}
\end{equation}
and
\begin{equation}
{w}_s(\mu,\mu',\vartheta')=\sigma_a\sqrt{\left(\frac 1 \mu+\frac h 2 \mu\vartheta^{\prime}\right)^{2}+\frac{h^{2} \mu^{\prime^{2}}}{4}}+\sigma_b\sqrt{\left(\frac 1 \mu-\frac h 2 \mu\vartheta^{\prime}\right)^{2}+\frac{h^{2} \mu^{\prime^{2}}}{4}}.
\end{equation}
The total energy is simply given by 
\begin{equation}\label{eq:1}
\mathcal E=\int_0^\ell w(\mu,\mu',\vartheta') {\rm d}x,
\end{equation}
where ${w}={w}_b+{w}_s$. Note that we have used, and from here on we use, the symbol $x$ to denote the variable $x_1$.

\subsection{Non-dimensionalization and approximation}\label{sec:non-dimens-appr}
We find it convenient to work with adimensional quantities.
For this reason, we introduce the following parameters
\begin{equation}\label{const}
\varepsilon=\frac h \ell,
\qquad 
\alpha=\frac{\alpha_{1}}{\alpha_{1}+\alpha_{2}},
\qquad 
\sigma=\frac 12 \frac{\sigma_{a}-\sigma_b}{\sigma_{a}+\sigma_{b}},
\qquad 
\gamma=\frac{\sigma_{a}+\sigma_{b}}{\ell\left(\alpha_{1}+\alpha_{2}\right)}.
\end{equation}
The parameters $\varepsilon, \alpha$, and $\gamma$ are non-negative. 
We will refer to $\sigma$ as the \textit{apico-basal imbalance parameter}, which will have a pivotal role in our analysis. Then \eqref{eq:1} becomes
\begin{equation}
\mathcal E=(\alpha_1+\alpha_2)\ell^2\int_0^1 \widetilde{w} d\widetilde x,
\end{equation}
where we have set
\begin{equation}
\tilde x=\frac x \ell,\qquad \tilde{w}=\widetilde{w}_b+\gamma\widetilde{w}_s,
\end{equation}
and where we have defined the dimensionless surface and bulk energies as, respectively,
\begin{equation}
\begin{aligned}
\widetilde{w}_s=&\frac{1}{2}\left(\sqrt{\left(\frac{1}{\tilde{\mu}}+\frac{{\varepsilon}}{2} \tilde{\mu} \tilde{\vartheta}^{\prime}\right)^{2}+\frac{{\varepsilon}^{2} \tilde{\mu}^{\prime 2}}{4}}+\sqrt{\left(\frac{1}{\tilde{\mu}}-\frac{{\varepsilon}}{2} \tilde{\mu} \tilde{\vartheta}^{\prime}\right)^{2}+\frac{{\varepsilon}^{2} \tilde{\mu}^{\prime 2}}{4}}\right) \\
&+\frac{\sigma}{2}\left(\sqrt{\left(\frac{1}{\tilde{\mu}}+\frac{{\varepsilon}}{2} \tilde{\mu} \tilde{\vartheta}^{\prime}\right)^{2}+\frac{{\varepsilon}^{2} \tilde{\mu}^{\prime 2}}{4}}-\sqrt{\left(\frac{1}{\tilde{\mu}}-\frac{{\varepsilon}}{2} \tilde{\mu} \tilde{\vartheta}^{\prime}\right)^{2}+\frac{{\varepsilon}^{2} \tilde{\mu}^{\prime 2}}{4}}\right)
\end{aligned}
\end{equation}
and
\begin{equation}
\widetilde{w}_b=\frac{\mathcal E_b}{(\alpha_1+\alpha_2)\ell}=
\varepsilon\left(\frac{1}{\tilde{\mu}^{2}}+\tilde{\mu}^{2}\right)+\frac{\varepsilon^{3}}{12}\left(\tilde\mu^{2} \tilde{\vartheta}^{\prime 2}+\alpha \frac{\tilde{\mu}^{\prime 2}}{2}\right).
\end{equation}
Note that $\tilde w=\frac{w_b+w_s}{\ell(\alpha_1+\alpha_2)}$. In what follows we drop tildas.

Since the bulk energy involves powers of $\varepsilon$ up to the third order, for consistency, we expand the surface energy up to the same order:
$$
w_{s} \simeq \frac{1}{\mu}+\frac{1}{2} \sigma \varepsilon \mu \vartheta^{\prime}+\frac{1}{8} \varepsilon^{2} \mu (\mu^{\prime})^2-\frac{1}{16} \sigma \varepsilon^{3} \mu^{3} \vartheta^{\prime} (\mu^{\prime})^2.
$$
The total energy that we consider hereafter is
\begin{equation}\label{wtot}
\begin{aligned}
w=& \, \varepsilon\left(\frac{1}{\mu^{2}}+\mu^{2}-(4-3\alpha)\left(\frac 1\mu+\mu\right)\right)+\frac{\varepsilon^{3}}{12}\left(\mu^{2} (\vartheta^{\prime})^2+\alpha \frac{(\mu^{\prime})^2}{2}\right)\\
&\quad + \gamma\left(\frac{1}{\mu}+\frac{1}{2} \sigma \varepsilon \mu \vartheta^{\prime}+\frac{1}{8} \varepsilon^{2} \mu (\mu^{\prime})^2-\frac{1}{16} \sigma \varepsilon^{3} \mu^{3} \vartheta^{\prime}(\mu^{\prime})^2\right).
\end{aligned}
\end{equation}

\subsection{Euler-Lagrange equations}
On denoting by $\partial_i w$ the partial derivative of $w$ with respect to the $i$-th argument, the Euler-Lagrange equations of the variational problem are
\begin{equation}\label{eq:2}
\begin{cases}
\begin{aligned}
&\partial_1w(\mu,\mu',\vartheta')-(\partial_2w(\mu,\mu',\vartheta'))'=0,\\
&(\partial_3w(\mu,\mu',\vartheta'))'=0,
\end{aligned}
\end{cases}
\end{equation}
while the natural boundary conditions are
\begin{equation}
\begin{cases}
\mu=\mu_\flat\qquad \text{or}\qquad \partial_2 w=0,\\
\vartheta=\vartheta_\flat\qquad\text{or}\qquad\partial_3w=0,
\end{cases}
\end{equation}
where $\mu_\flat$ and $\vartheta_\flat$ are given.

Before proceeding, we find it worth observing that the system of Euler-Lagrange equations is autonomous; as a result, we can transform it into a first-order system by introducing two  integration constants. Although we do not make use of this observation in the present paper, it might prove useful for further work, and therefore we sketch the derivation. The second equation of \eqref{eq:2} can be integrated to give:
\begin{equation}\label{M}
\partial_{3} w\left(\mu, \mu^{\prime}, \vartheta^{\prime}\right)=M,
\end{equation}
where $M$ is a constant. 
Also, noticing that
\begin{equation}
\begin{aligned}
\left(w-\mu^\prime \partial_2w\right)'&=
\partial_1w \mu'+\partial_2w\mu''+\partial_3w\vartheta''-\mu'' \partial_2w-\mu^\prime \left(\partial_2w\right)'
\\
&   =   \left( \partial_1w-\left(\partial_2w\right)' \right)\mu'+\partial_3w\vartheta''\\
&   =   M\vartheta'',
\end{aligned}
\end{equation}    
where the last equality has been obtained by using \eqref{eq:2} and \eqref{M}, we deduce that
there is a constant $C$ such that
\begin{equation}\label{C}
w\left(\mu, \mu^{\prime}, \vartheta^{\prime}\right)-\mu^{\prime} \partial_{2} w\left(\mu, \mu^{\prime}, \vartheta^{\prime}\right)=M \vartheta^{\prime}+C.
\end{equation}
Equations \eqref{M} and \eqref{C} can be used in place of \eqref{eq:2}.

\subsection{The uniformly straight configuration}\label{straight}
By a uniformly straight configuration we mean a configuration of the form
\begin{equation}\label{eq:5}
\mu(x)=\bar\mu_0,\quad\vartheta(x)=\bar\vartheta_0,   
\end{equation}
with $\bar\mu_0$ and $\bar\vartheta_0$ constant.
Since the Euler-Lagrange equations are invariant under rigid rotations, we may take, without loss of generality, 
$\bar\vartheta_0=0.$

\begin{figure}[h]
	\centering
	\includegraphics[scale=.5]{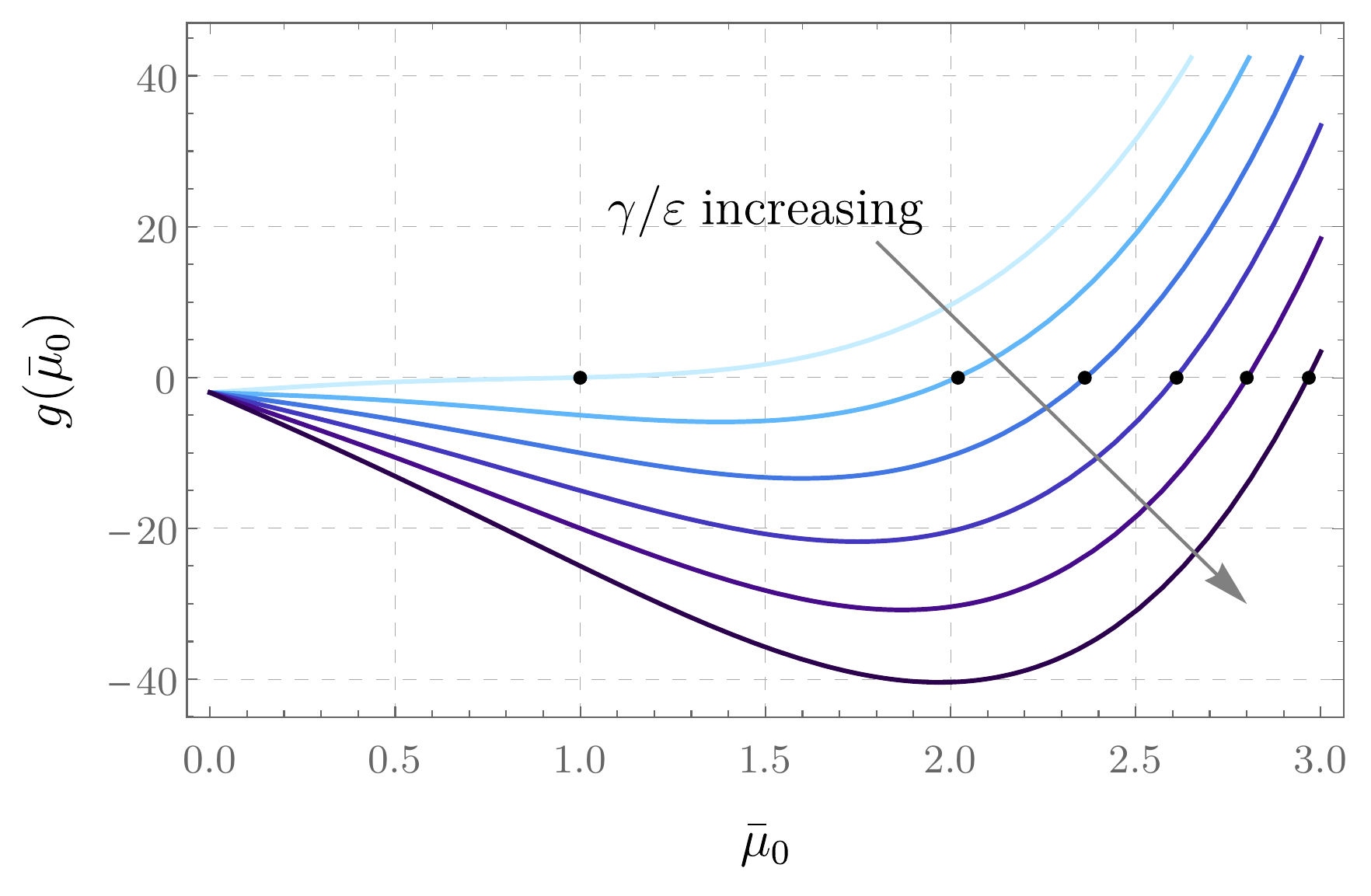}
	\caption{Function $g(\bar{\mu}_0)$ in \eqref{mu0},  for fixed $\alpha=0.2$. The points where $g(\bar{\mu}_0)$ vanishes correspond to the value of $\mu$ in the straight configuration. We notice that, when the surface energy has a more and more prevalent role, the thickness of body increases.}
	\label{fig:gmu0}
\end{figure}

We now show that for every value of the parameters $\varepsilon, \alpha,\sigma$, and $\gamma$ there is a uniformly straight configuration, characterized by a particular value of $\bar\mu_0$, that satisfies the Euler-Lagrange equations.
Indeed, this is quite simple to check by means of \eqref{eq:2}. In fact, since $\partial_2w(\bar\mu_0,0,0)$ and $\partial_3w(\bar\mu_0,0,0)$ are constant, the equations \eqref{eq:2} simply reduce to
\begin{equation}\label{mu00}
\partial_1w(\bar\mu_0,0,0)=0.
\end{equation}
This equation can be used to find $\bar\mu_0$. By evaluating
\begin{equation}\label{eq:7}
w(\bar\mu_0,0,0)=\frac{\varepsilon+\gamma \bar\mu_0+\varepsilon \big(\bar\mu_0^4-(4-3\alpha)(\bar\mu_0+\bar\mu_0^3)\big)}{\bar\mu_0^{2}},
\end{equation}
we find
\begin{equation}
\partial_1w(\bar\mu_0,0,0)=\frac{ \left(\bar\mu_0^2-1\right)\big( 2-\bar\mu_0(4-3\alpha-2\bar\mu_0)\big) \varepsilon -\gamma \bar\mu_0}{\bar\mu_0^3}.
\end{equation}
Equation \eqref{mu00} leads to
\begin{equation}\label{mu0}
g(\bar\mu_0):=\bar\mu_0^4  -\frac{\gamma}{2\varepsilon} \bar\mu_0-\frac{4-3\alpha}{2} \bar\mu_0( \bar\mu_0^2-1)-1=0,
\end{equation}
and it is easily concluded that $g$ has only one positive root (see. Fig. \ref{fig:gmu0}), hereafter denoted by $\mu_0$. 

Moreover, since $g(1)=-\gamma/(2\varepsilon)\le 0$,
we see that
\begin{equation}\label{mu0>1}
\mu_0\ge 1 \quad \mbox{and}\quad \mu_0=1 \iff \gamma=0.
\end{equation}
Thus, if the surface energy is not ``activated'' ($\gamma=0$) there is no stretching of the transversal sections $\mu_0=1$ and by the incompressibility constraint, \eqref{lm1}, also the axial stretching is absent. This is clearly consistent with the assumption of natural reference configuration.
While, if the surface energy is ``activated'' ($\gamma>0$), the surface area is penalized and hence the axial stretching decreases. By the incompressibility constraint it follows that the transversal sections are stretched: $\mu_0>1$.



\subsection{The boundary value problem}

We  consider the following boundary conditions:
$$
\mu(x)=\mu_{0}, \quad \vartheta(x)=0 \quad \text { for } \quad x=0,1,
$$
where $\mu_{0}$ is the root of $(37) .$ 
From Section \ref{straight} we know that configurations of the form
$$
\mu(x)=\mu_{0}, \quad \vartheta(x)=\vartheta_{0}
$$
satisfy the equilibrium equations.

As detailed in the Introduction, experimental evidences show that imbalance between apical and basal surface tension may induce a change of shape in cancerous epithelial tissues; thus, we investigate the existence of other (non trivial) solutions as the apico-basal imbalance parameter $\sigma$ changes. It is convenient to set
$$
\xi(x)=\mu(x)-\mu_{0}(x)
$$
and to denote by
$$
\uu=(\xi, \vartheta)
$$
the pair of independent variables. We shall explicitly write the dependence of the energy on $\sigma$
$$
\mathcal{E}(\uu, \sigma)=\int_{0}^{1} w\left(\mu_{0}+\xi, \xi^{\prime}, \vartheta^{\prime}\right) d x
$$
where $w$ is given by \eqref{wtot}. The equilibrium equations are
\begin{equation}\label{eqb1}
0=d \mathcal{E}(\uu, \sigma)[\vv]=\left.\frac{d}{d s} \mathcal{E}(\uu+s \vv, \sigma)\right|_{s=0} \quad \forall \vv;
\end{equation}
every time we write $\forall \vv$, we mean for every $\xi$, $\vartheta$ smooth that are null at $0$ and $1$.
This equation, after integration by parts and localization, is equivalent to \eqref{eq:2}.
Since the uniformly straight configuration $\xi=0$ and $\vartheta=0$ satisfies the equilibrium condition for every $\sigma$, we have that
\begin{equation}\label{trivial}
d \mathcal{E}(\oo, \sigma)[\vv]=0 \quad \forall \vv \text { and } \forall \sigma
\end{equation}
where we have set
$$
\oo=(0,0).
$$

\section{Critical apico-basal imbalance: loss of positivity of the elasticity tensor}

As reported in the Introduction, in  \cite{Messal} it has been observed, by detecting 
the intensity of pMLC2 staining on the apical and the basal side of an epithelium, that the difference
between the apical and basal surface tension is considerably higher in wild-type cells than in transformed cells. 
This means that the parameter $\sigma$, which measures the surface tension apico-basal imbalance, in transformed cells is small compared to the value assumed by $\sigma$ in wild-type cells.
From this consideration a natural question arises: do the sole change of  $\sigma$ leads to a morphological transition? 
From  Section \ref{straight} we know that for every value of $\sigma$ there is a uniformly straight configuration and hence if a morphological transformation occurs we will have more than one solution.
This occurrence is not possible if the elasticity tensor (\textit{i.e.} the second variation of the energy) is always positive definite. 

In this section we determine the values of $\sigma$ for which there could be a second solution in a neighborhood  of the uniformly straight configuration ($\xi=0$ and $\vartheta=0$). For these critical values not only the elasticity tensor loses its positive definiteness but also the assumptions of the implicit function theorem fail. This analysis will open the stage to the bifurcation analysis carried on in the next section.

By the implicit function theorem, a necessary condition for $\sigma=\sigma_{c}$ to be a bifurcation point is
\begin{equation}\label{uc}
d^{2} \mathcal{E}\left(\oo, \sigma_{c}\right)\left[\uu_{c}, \vv\right]=0 \quad \forall \vv,
\end{equation}
for some non trivial $\uu_{c}=\left(\xi_{c}, \vartheta_{c}\right)$ satisfying the boundary conditions, where
$$
d^{2} \mathcal{E}(\uu, \sigma)[\vv, \ww]=\left.\frac{d}{d t} d \mathcal{E}(\uu+t \ww, \sigma)[\vv]\right|_{t=0}=\left.\frac{d^{2}}{d s d t} \mathcal{E}(\uu+s \vv+t \ww, \sigma)\right|_{s, t=0}.
$$

The second derivative of the energy can be written explicitly as
\begin{equation}
\begin{aligned}
d^{2} \mathcal{E}(\uu, \sigma)[\hat{\vv}, \overline{\vv}]=\int_{0}^{1} & \partial_{11}^{2} w \hat{\xi} \bar{\xi}+\partial_{12}^{2} w\left(\hat{\xi}^{\prime} \bar{\xi}+\hat{\xi} \bar{\xi}^{\prime}\right)+\partial_{13}^{2} w\left(\hat{\xi} \bar{\vartheta}^{\prime}+\hat{\vartheta}^{\prime} \bar{\xi}\right) \\
&+\partial_{22}^{2} w \hat{\xi}^{\prime} \bar{\xi}^{\prime}+\partial_{23}^{2} w\left(\hat{\xi}^{\prime} \bar{\vartheta}^{\prime}+\hat{\vartheta}^{\prime} \bar{\xi}^{\prime}\right)+\partial_{33}^{2} w \hat{\vartheta} \bar{\vartheta}^{\prime} d x
\end{aligned}
\end{equation}
and evaluating all the second derivatives of $w$ at $\left(\mu, \mu^{\prime}, \vartheta\right)=\left(\mu_{0}, 0,0\right)$ we find
\begin{equation}\label{d2E}
d^{2} \mathcal{E}\left(\oo, \sigma_{c}\right)[\hat{\vv}, \overline{\vv}]=\int_{0}^{1} \CC_{11} \hat{\xi} \bar{\xi}+\sigma_{c} \CC_{13}\left(\hat{\xi} \bar{\vartheta}^{\prime}+\hat{\vartheta}^{\prime} \bar{\xi}\right)+\CC_{22} \hat{\xi}^{\prime} \bar{\xi}^{\prime}+\CC_{33} \hat{\vartheta}^{\prime} \bar{\vartheta}^{\prime} d x
\end{equation}
where
\begin{equation}\label{CC}
\begin{array}{ll}
\displaystyle\CC_{11}=\partial_{11}^{2} w\left(\mu_{0}, 0,0\right)=2\left(\varepsilon+\frac{3 \varepsilon+\big(\gamma-(4-3\alpha)\varepsilon \big)\mu_{0}}{\mu_{0}^{4}}\right),& \displaystyle\CC_{13}=\frac{\partial_{13}^{2} w\left(\mu_{0}, 0,0\right)}{\sigma}=\frac{\varepsilon \gamma}{2}, \\
\displaystyle\CC_{22}=\partial_{22}^{2} w\left(\mu_{0}, 0,0\right)=\frac{\varepsilon^{2}}{12}\left(\alpha \varepsilon+3 \gamma \mu_{0}\right), &\displaystyle \CC_{33}=\partial_{33}^{2} w\left(\mu_{0}, 0,0\right)=\frac{\varepsilon^{3} \mu_{0}^2}{6}.
\end{array}
\end{equation}
Then
$$
\begin{aligned}
d^{2} \mathcal{E}\left(\oo, \sigma_{c}\right)\left[\uu_{c}, \vv\right] &=\int_{0}^{1} \CC_{11} \xi_{c} \xi+\sigma_{c} \CC_{13}\left(\xi_{c} \vartheta^{\prime}+\vartheta_c^{ \prime} \xi\right)+\CC_{22} \xi^{c{\prime}} \xi^{\prime}+\CC_{33} \vartheta_c^{ \prime} \vartheta^{\prime} d x \\
&=\int_{0}^{1}\left(\CC_{11} \xi_{c}+\sigma_{c} \CC_{13} \vartheta_c^{ \prime}-\CC_{22} \xi_c^{ \prime \prime}\right) \xi-\left(\sigma_{c} \CC_{13} \xi^{c{\prime}}+\CC_{33} \vartheta_c^{ \prime \prime}\right) \vartheta d x
\end{aligned}
$$
where we have taken into account that $\uu_{c}$ and $\vv=(\xi, \vartheta)$ are null at $x=0,1$. Condition \eqref{uc} is therefore equivalent to the following system of differential equations:
\begin{equation}\label{sysc}
\begin{cases}
\CC_{11} \xi_{c}+\sigma_{c} \CC_{13} \vartheta_c^{\prime}-\CC_{22} \xi_c^{ \prime \prime}=0, \\
\sigma_{c} \CC_{13} {\xi_{c}}'+\CC_{33} \vartheta_c^{ \prime \prime}=0.
\end{cases}
\end{equation}
For $\sigma_c \neq 0$, since $\CC_{13} \neq 0$, it is possible to get rid off ${\xi_{c}}''$ in \eqref{sysc}$_{1}$ by determining ${\xi_{c}}'$ from \eqref{sysc}$_{2}$ and then solve for   ${\xi_{c}}$: 
\begin{equation}\label{xicd}
\xi_{c}=-\frac{1}{\CC_{11}}\left(\sigma_{c} \CC_{13} \vartheta_c^{ \prime}+\frac{\CC_{22} \CC_{33}}{\sigma_{c} \CC_{13}} \vartheta_c^{ \prime \prime\prime}\right),
\end{equation}
where $\vartheta_{c}$ is the solution of  the following problem:
$$
\begin{cases}
\CC_{22} \CC_{33} \vartheta_c^{ \prime \prime \prime \prime}+\left(\sigma_c^{2} \CC_{13}^{2}-\CC_{11} \CC_{33}\right) \vartheta_c^{ \prime \prime}=0 \quad \text { in }(0,1), \\
\vartheta_{c}=0 \qquad\text { at } x=0,1, \\
\sigma_{c} \CC_{13} \vartheta_c^{^{\prime}}+\dfrac{\CC_{22} \CC_{33}}{\sigma_{c} \CC_{13}} \vartheta_c^{ \prime \prime \prime}=0 \qquad \text { at } x=0,1.
\end{cases}
$$
For $\sigma_{c} \CC_{13}^{2}-\CC_{11} \CC_{33} \leq 0$ the only solution is $\vartheta_{c}=0$. For $\sigma_{c} \CC_{13}^{2}-\CC_{11} \CC_{33}>
0$ set
$$
\omega^{2}=\frac{\sigma_c^{2} \CC_{13}^{2}-\CC_{11} \CC_{33}}{\CC_{22} \CC_{33}},
$$
and write the general solution as
\begin{equation}\label{tcd}
\vartheta_{c}(x)=a_{1} \cos (\omega x)+a_{2} \sin (\omega x)+a_{3} x+a_{4}.
\end{equation}
Imposing the boundary conditions we find the system
\begin{equation}\label{ai}
\left(\begin{array}{cccc}
1 & 0 & 0 & 1 \\
\cos \omega & \sin \omega & 1 & 1 \\
0 & \CC_{11} \CC_{33} \omega & \sigma_c^{2} \CC_{13}^{2} & 0 \\
-\CC_{11} \CC_{33} \omega \sin \omega & \CC_{11} \CC_{33} \omega \cos \omega & \sigma_c^{2} \CC_{13}^{2} & 0
\end{array}\right)\left(\begin{array}{l}
a_{1} \\
a_{2} \\
a_{3} \\
a_{4}
\end{array}\right)=0,
\end{equation}
from which we deduce that we have a non trivial solution if and only if
$$
2\sigma_c^{2} \CC_{13}^{2}(1-\cos \omega)=\CC_{11} \CC_{33} \omega \sin \omega.
$$
This equation is satisfied if
\begin{equation}\label{omega}
\omega=2 k \pi \quad \text { for } k=1,2,3, \ldots
\end{equation}
while for $\omega \neq 2 k \pi$ we may divide the equation by $\sin \omega,$ after noticing that $\omega=(2 k+1) \pi$ is not a solution, to find
\begin{equation}\label{secsol}
\tan \frac{\omega}{2}=\frac{\CC_{11} \CC_{33}}{2\sigma_c^{2} \CC_{13}^{2}} \omega=\frac{(\omega / 2)\left(\CC_{11} / \CC_{22}\right)}{\omega^{2}+\CC_{11} / \CC_{22}},
\end{equation}
where we used the equality $\tan (z / 2)=(1-\cos z) / \sin z$.
It is easy to see that this equation has an infinite number of solutions. We denote the smallest solution different from zero by $\hat{\omega}$ and we remark that $\hat{\omega}>2 \pi,$ as an analytical study of the functions involved show. 
\begin{figure}[h!]
	\centering
	\includegraphics[scale=.5]{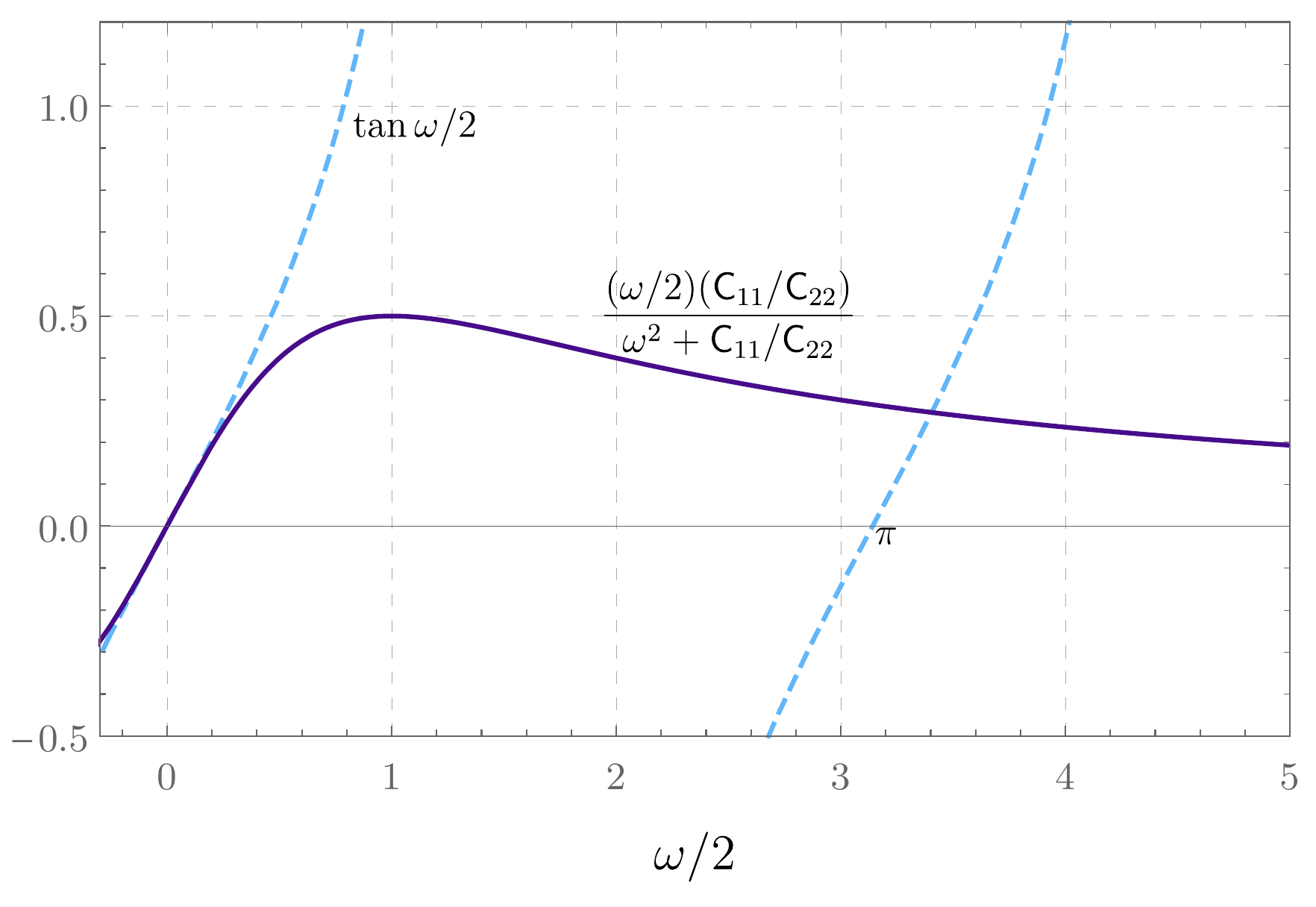}
	\caption{A graphical solution of equation \eqref{secsol}. The left hand-side of \eqref{secsol} is depicted with a dashed line while the right hand-side with a solid line. The smallest point, larger than zero, in which the two graphs intersect is for $\omega/2>\pi$.}
	\label{fig:caricocritico}
\end{figure}

The smallest value of $\omega$ for which we might have a non trivial solution, and hence a bifurcation, is given by \eqref{omega} for $k=1$:
$$
\omega_{c}=2 \pi.
$$
From the definition of $\omega$ we deduce the smallest value of $\sigma_{c}$ for which we might have a bifurcation
$$
\omega_{c}^{2}=4 \pi^{2}=\frac{\sigma_c^{2} \CC_{13}^{2}-\CC_{11} \CC_{33}}{\CC_{22} \CC_{33}},
$$
that is
\begin{equation}\label{sigmac}
\sigma_{c}=\pm \sqrt{\frac{4 \pi^{2} \CC_{22}+\CC_{11}}{\CC_{13}^{2}} \CC_{33}}.
\end{equation}
On recalling  definitions  \eqref{CC}, we can express  the critical value of the apico-basal imbalance parameter as
\begin{equation}
\sigma_{c}=\pm\frac{\sqrt{2}\gamma}{\sqrt{3\varepsilon}\mu_0}\sqrt{2\varepsilon\left( 1+\frac{3}{\mu_0^4} +\frac{\gamma-(4-3\alpha)\varepsilon}{\mu_0^3}\right)+4\pi^2\left(  \frac{\alpha\varepsilon^3}{12}+\frac 14 \gamma\varepsilon^2\mu_0\right) }.
\end{equation}
Since the parameters  $\gamma$, $\varepsilon$, $\alpha$ and $\mu_0$ are related by \eqref{mu0},
we may express $\sigma_c$ in terms of only three variables. We find, \textit{e.g.},
\begin{equation}
\begin{aligned}
\sigma_{c}=&\pm\frac{}{}\frac{\sqrt{2}\mu_0^2}{\sqrt{3\varepsilon}(2-(4-3\alpha)(\mu_0-\mu_0^3)-\mu_0^4)}\times\\
&\sqrt{\frac{2\varepsilon}{\mu_0^4}\big(1-(4-3\alpha-3\mu_0)\mu_0^3\big)+\frac{\pi^2\varepsilon^3}{3}\big( 6(\mu_0-1)^3(\mu_0+1)+\alpha(1-9(\mu_{0}-\mu_0^3))  \big)}.
\end{aligned}
\end{equation}
It is also possible to express $\sigma_c$ in terms of $\gamma$, $\varepsilon$, and $\alpha$ by solving \eqref{mu0} in terms of $\mu_0$. We refrain to write this expression of $\sigma_c$ since it would involve cumbersome quartic roots of \eqref{mu0}.

\begin{figure}[h!]
	\centering
	\includegraphics[scale=.45]{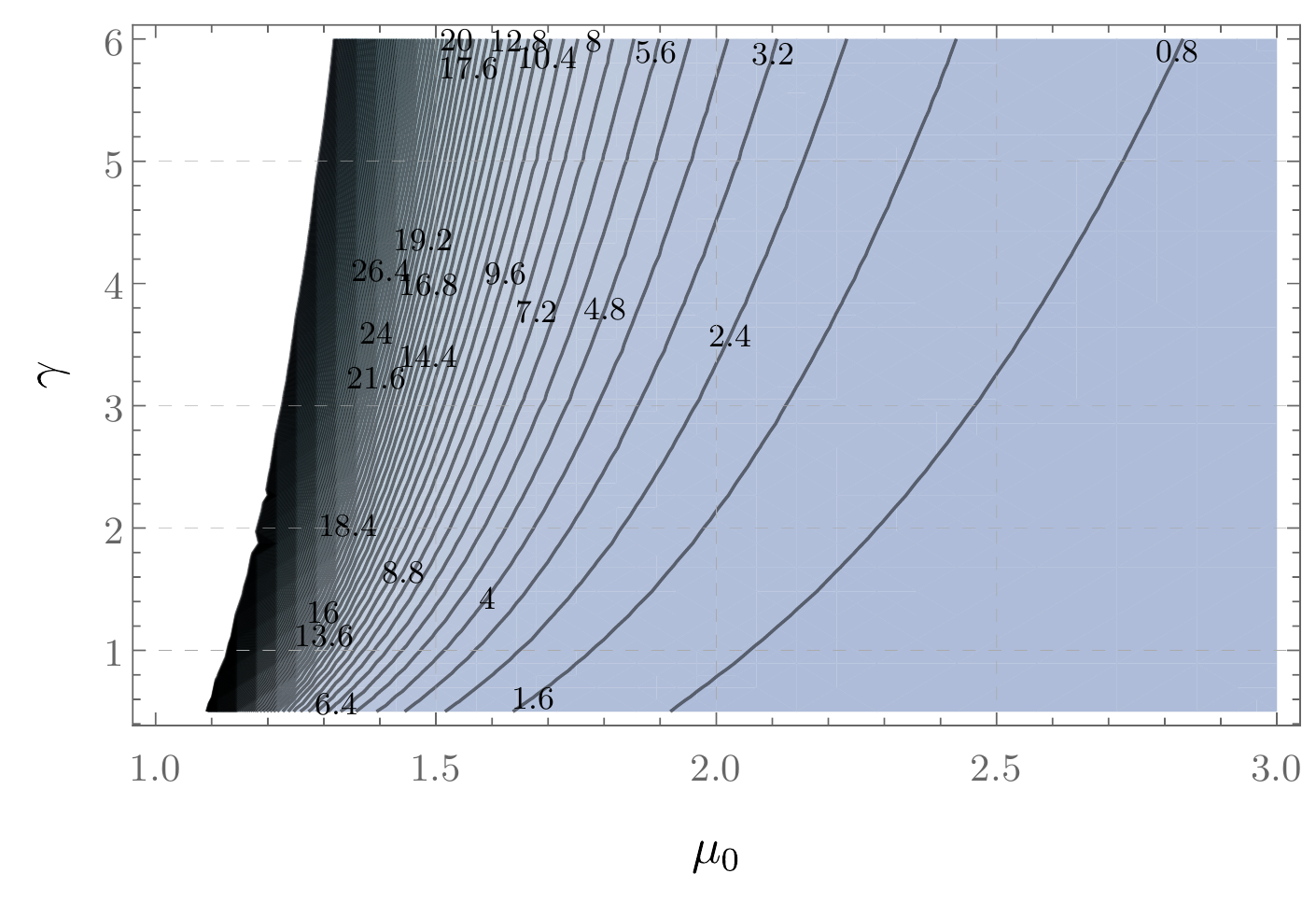}
	\includegraphics[scale=.45]{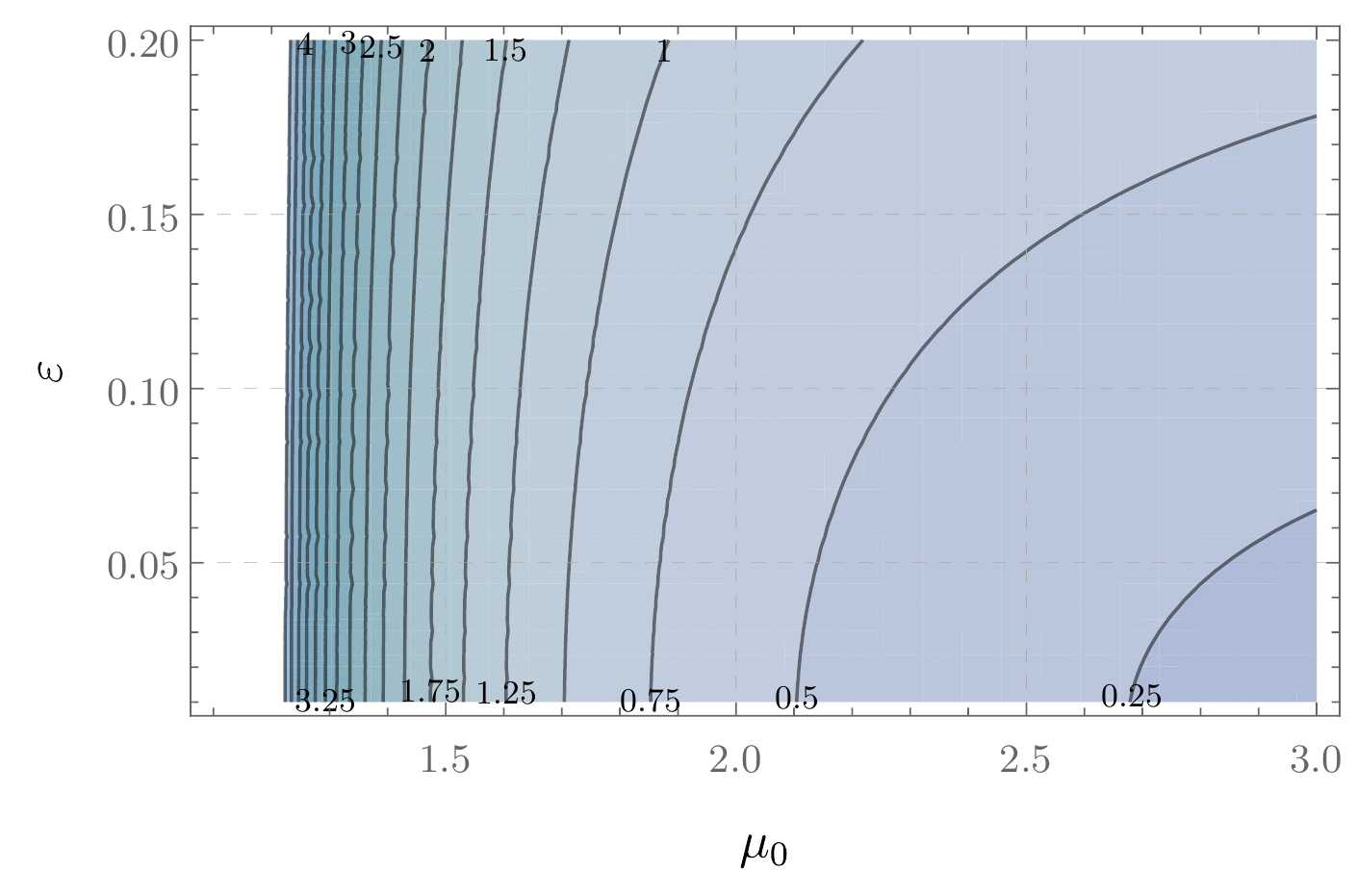}
	\includegraphics[scale=.45]{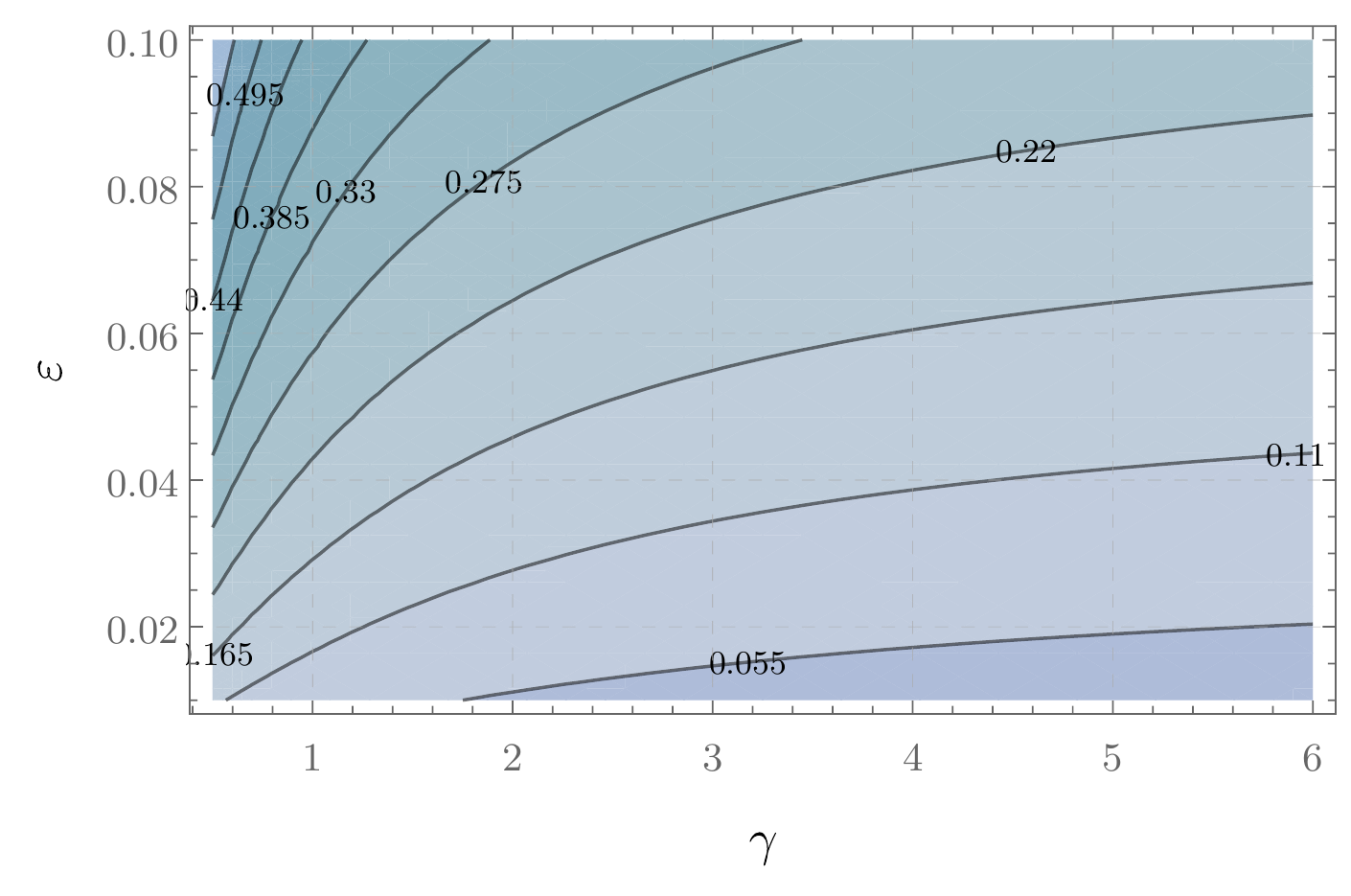}
	\caption{The contour plots for the critical apico-basal imbalance parameter $\sigma_c$, for $\alpha=0.2$, and as a function of $\gamma$ and $\mu_0$ (top left), $\varepsilon$ and $\mu_0$ (top right), $\varepsilon$ and $\gamma$ (center below). The possible values that $\sigma$ can take are between $0$ and $0.5$, thus in the regions  in which $\sigma_c$ is larger than $0.5$  no bifurcation occurs. }
	\label{fig:sigmac}
\end{figure}

In Fig. \ref{fig:sigmac} we plot $\sigma_c$, for fixed $\alpha$ and $\varepsilon$, as a function of $\gamma$ and $\mu_0$; in Table \ref{tablee} we report the values of $\sigma_{c}$ relative to $\varepsilon=0.1$, for several values of $\alpha$ and $\gamma$. The possible values that $\sigma$ can take are between $0$ and $0.5$, thus for $\sigma_c >0.5$  no bifurcation occurs. The competition between bulk energy and surface energy is captured by $\gamma$. In Table  \ref{tablee} we see that for $\gamma$ small the stabilizing effect of the  bulk energy prevails and the bifurcation is inhibited.

%

\begin{table}
	\begin{tabular}{cccccccccc}
		\toprule
		&&	 \multicolumn{8}{c}{$\gamma$} \\
		& &0.3&0.5&1 &2& 3 & 4& 5& 6\\
		\cmidrule{3-10}
		\multirowcell{6}{$\alpha$} &0.0\hspace{.3cm}\vline&	\cellcolor{Gray}0.8973 & \cellcolor{Gray}0.6743 & 0.4794 & 0.3649& 0.3219 & 0.2988 & 0.2841 & 0.2738 \\
		&0.2\hspace{.3cm}\vline&\cellcolor{Gray}0.8478 &\cellcolor{Gray} 0.6385 & 0.4542 & 0.346 & 0.3056 & 0.2842 & 0.2707 & 0.2613 \\
		&0.4\hspace{.3cm}\vline&\cellcolor{Gray}	0.8299 & \cellcolor{Gray}0.618 & 0.4355 & 0.3301 & 0.2915 & 0.2713 & 0.2587 & 0.2500 \\
		&0.6\hspace{.3cm}\vline&\cellcolor{Gray}	0.8365 &\cellcolor{Gray} 0.6102 & 0.4225 & 0.3171 & 0.2793 & 0.2598 & 0.2479 & 0.2398\\
		&0.8\hspace{.3cm}\vline&\cellcolor{Gray}	0.8590 & \cellcolor{Gray}0.6121 & 0.4143 & 0.3066 & 0.2689 & 0.2498 & 0.2383 & 0.2306 \\
		&1.0\hspace{.3cm}\vline&\cellcolor{Gray}0.8905&\cellcolor{Gray} 0.6208& 0.4100 & 0.2983 & 0.2601 & 0.2411 & 0.2297 & 0.2222\\
		\bottomrule
	\end{tabular}
	\caption{Values of $\sigma_{c}$ relative to $\varepsilon=0.1$, for several values of $\alpha$ and $\gamma$. Gray cells correspond to non admissible values for $\sigma_c$.}
	\label{tablee}
\end{table}

From the system \eqref{ai}, with $\omega=\omega_{c}$, we find $a_{2}=a_{3}=0$ and $a_{1}=-a_{4},$ and from \eqref{tcd} and \eqref{xicd} we deduce, up to a constant,
$$
\vartheta_{c}(x)=\K(1-\cos (2 \pi x)), \qquad \xi_{c}=\sin (2 \pi x),
$$
where the constant $\kappa$ is defined by
\begin{equation}\label{K}
\K=-\frac{\sigma_c\CC_{13}}{2\pi\CC_{33}}.
\end{equation}
Thus the eigenspace associated to $\sigma_{c}$, given by \eqref{sigmac}, is one dimensional:
\begin{equation}\label{Nuc}
\NN=\operatorname{span}\left\{\uu_{c}\right\}, \quad \mbox{with}\quad\uu_{c}=(\xi_c,\vartheta_c)=(\sin (2 \pi x), \K(1-\cos (2 \pi x))).
\end{equation}

\begin{figure}[h!]
	\centering
	\includegraphics[scale=.45]{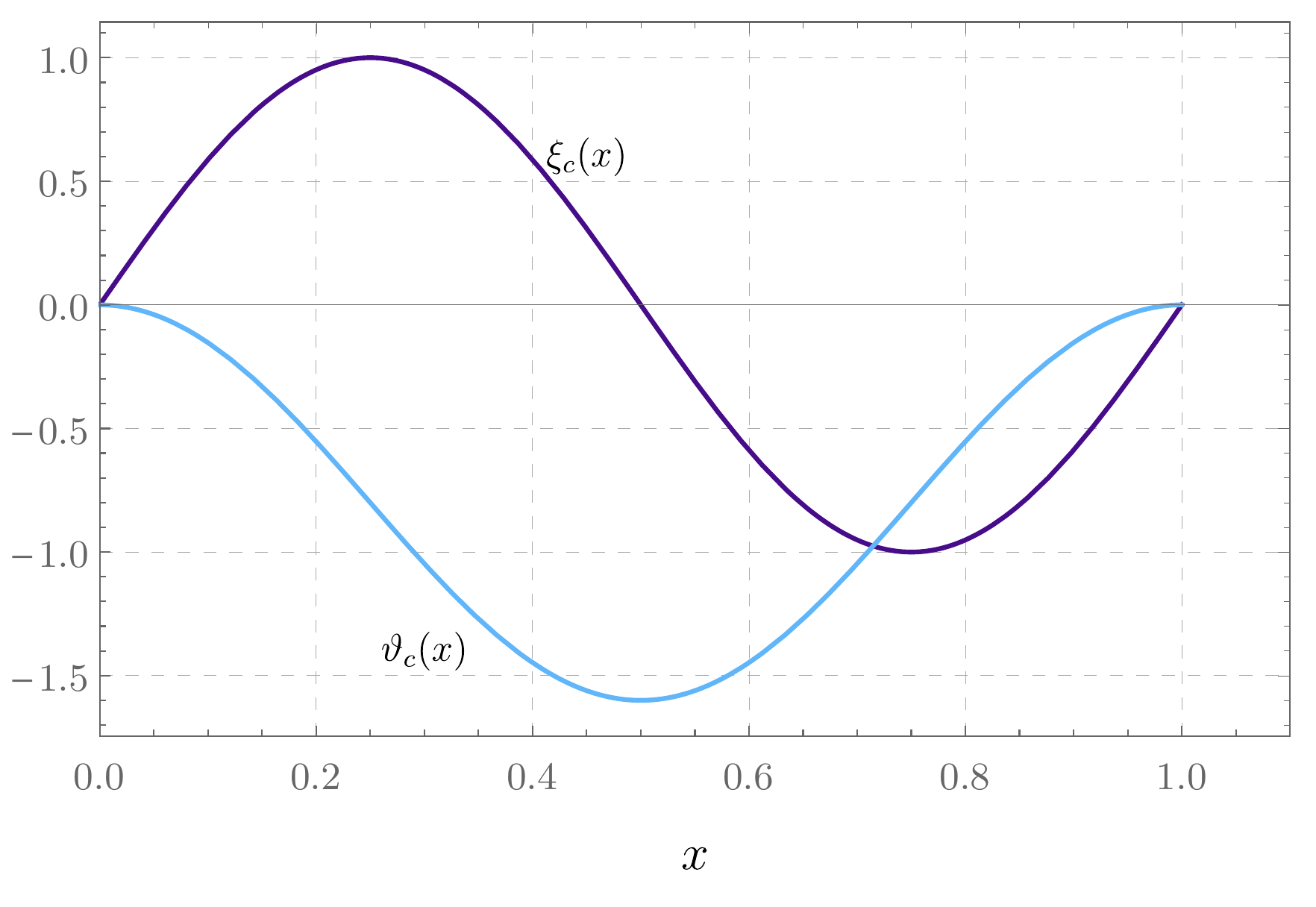}
	\caption{The critical mode $\uu_c$.}
	\label{fig:modes}
\end{figure}

\section{Bifurcation analysis}
In the previous section we found the smallest value $\sigma_c$ of the surface tension apico-basal imbalance for which a bifurcation may occur. 
In this section, by means of the  \textit{Lyapunov-Schmidt decomposition method}, we show that the critical value $\sigma_c$ is indeed a bifurcation point, hence proving that besides the uniformly straight configuration there is a ``folded'' configuration. For $\sigma$ less than the critical value $\sigma_c$, the stabilizing contribution of the epithelial elasticity keeps the equilibrium configuration straight, while for $\sigma$ lager than $\sigma_c$ the surface energy prevails and the epithelia bifurcates. In Section \ref{sub41} we characterize the bifurcation and we make further remarks.

Within the framework of the Lyapunov-Schmidt decomposition, the increment $\uu-\uu_c$ associated  to an increment of the parameter $\sigma$ is decomposed in two components: one is in $\NN$, and the other in the complement set
$$
\NN^\perp=\{\tilde \uu=(\tilde\xi,\tilde\vartheta): \int_0^1\uu_c\cdot\tilde \uu\,{\rm d}x= \int_0^1\xi_c\tilde\xi+\vartheta_c\tilde\vartheta\,{\rm d}x=0\}.
$$
A generic $\uu$ and $\sigma$ can be written uniquely as
\begin{equation}\label{decomp}
\begin{aligned}
& \uu=\eta\uu_c+\tilde\uu\quad \mbox{for some }\eta\in\mathbb{R} \mbox{ and }\tilde\uu\in\NN^\perp,\\
&\sigma=\sigma_c + \beta\quad\mbox{for some }\beta\in\mathbb{R}.
\end{aligned}
\end{equation}
We now decompose the test function $\vv$ appearing in \eqref{eqb1}  as in \eqref{decomp} and then we find
\begin{equation}\label{eqb2}
\begin{cases}
d\Ec(\eta\uu_c+\tilde\uu,\sigma_c +\beta)[\uu_c]=0,\\
d\Ec(\eta\uu_c+\tilde \uu,\sigma_c +\beta)[\tilde\vv]=0\quad \forall\tilde \vv\in \NN^\perp.
\end{cases}
\end{equation}

For fixed values of $\eta,\beta\in\mathbb{R}$ the second equation of \eqref{eqb2} can be uniquely solved for the unknown $\tilde \uu$. The solution
\begin{equation}\label{vsol}
\tilde \uu=\tilde \uu(\eta,\beta)
\end{equation}
satisfies
\begin{equation}\label{eqb3}
d\Ec(\eta\uu_c+\tilde \uu(\eta,\beta),\sigma_c +\beta)[\tilde\vv]=0\quad \forall\tilde \vv\in \NN^\perp\mbox{ and }\forall\eta,\beta\in\mathbb{R}
\end{equation}
and, in agreement with \eqref{trivial},
\begin{equation}\label{eqb4}
\tilde \uu(0,\beta)=0\quad \forall\beta\in\mathbb{R}.
\end{equation}
Let us now introduce the \textit{reduced function}
\begin{equation}\label{deff}
f(\eta,\beta)=d\Ec(\eta\uu_c+\tilde \uu(\eta,\beta),\sigma_c +\beta)[\uu_c];
\end{equation}
with this definition, solving  \eqref{eqb2}$_1$ reduces to  searching the roots of the reduced function.  By using \eqref{trivial} and \eqref{eqb4}, we first notice that
\begin{equation}\label{trivialf}
f(0,\beta)=d\Ec(0,\sigma_c +\beta)[\uu_c]=0\quad \forall \beta;
\end{equation}
hence $\eta=0$ is a root for every $\beta$.

We now look for roots different from the trivial solution $\eta=0$; more specifically, we plan to determine whether the equation $f(\eta,\beta)=0$  has a branch of solutions crossing the trivial one at $(0,\beta)$.

We first notice that the reduced function $f$ satisfies the following identities
\begin{equation}\label{propf}
f(0,0)=\frac{\partial f}{\partial\beta}(0,0)=\frac{\partial^2 f}{\partial\beta^2}(0,0)=\frac{\partial f}{\partial\eta}(0,0)=0
\end{equation}
and
$$
\frac{\partial^2 f}{\partial\eta\partial\beta}(0,0)=d^2\Ec_\sigma(0,\sigma_c)[\uu_c,\uu_c], \quad\mbox{with } \Ec_\sigma(\cdot,\sigma)=\frac{d}{d\sigma}\Ec(\cdot,\sigma).
$$
as shown in  \ref{appf}.
From \eqref{wtot} we easily get
$$
w_\sigma=\frac{\partial w}{\partial\sigma}= 
\gamma\left(\frac{1}{2}  \varepsilon \mu \vartheta^{\prime}-\frac{1}{16}  \varepsilon^{3} \mu^{3} \vartheta^{\prime} (\mu^{\prime})^{2}\right),
$$
and then we deduce that
$$
\partial_{13}^{2} w_\sigma\left(\mu_{0}, 0,0\right)= \frac{\gamma\varepsilon}{2}  \quad\text {and}\quad \partial_{ij}^{2} w_\sigma\left(\mu_{0}, 0,0\right)=0 \mbox{ for }(i,j)\notin\{(1,3),(3,1)\}.$$

Similarly to what we have already done in deducing  \eqref{d2E}, we find 
$$
\begin{aligned}
d^2\Ec_\sigma(\oo,\sigma_c)[\uu_c,\uu_c]&=\int_0^1 2\partial_{13}^{2} w_\sigma\left(\mu_{0}, 0,0\right)\xi_c\vartheta_c^{\prime}\,{\rm d}x\\
&=2\pi\gamma\varepsilon\K
\int_0^1 \sin^2(2\pi x)\,{\rm d}x=\pi\gamma\varepsilon\K.
\end{aligned}
$$

Hence
\begin{equation}\label{fab}
\frac{\partial^2 f}{\partial\eta\partial\beta}(0,0)=d^2\Ec_\sigma(0,\sigma_c)[\uu_c,\uu_c]=\pi\gamma\varepsilon\K\ne 0.
\end{equation}
Through a clear use of the implicit function theorem, equation \eqref{fab}, together with \eqref{propf}, allows to conclude  that there is a branch of solutions different from the trivial one through the point $(\eta,\beta)=(0,0)$. More precisely,  there exists a unique
$$
\beta=\beta(\eta) \quad\mbox{with}\quad  \beta(0)=0,
$$
such that
\begin{equation}\label{fb}
f(\eta,\beta(\eta))=0.
\end{equation}
In particular, this implies that $\sigma_c$ is a \textit{bifurcation point}.

\subsection{Apico-basal tension imbalance yields subcritical bifurcation}
\label{sub41}
When a  homogeneous  material is subjected to an elastic compression exceeding a certain threshold, wrinkling may occur  on the surface,  a bifurcation that usually is \textit{supercritical}. However, there are some circumstances in which bifurcations become \textit{subcritical}, \textit{e.g.}: (i) when the material is free-standing \cite{Hohlfeld2012}; (ii) for a bilayer with the same elastic moduli for substrate and coating,
but with a compressed substrate \cite{Chen2014}; (iii) if there is a  large enough surface energy \cite{Chen2012}.

Hereafter we show that, according to our model,  the competition between epithelial tissue elasticity and surface tension produces a subcritical bifurcation, when the critical value $\sigma_c$ of the apico-basal imbalance parameter is reached. 
This, in particular, implies that the uniformly straight configuration is stable for $\sigma$ below  $\sigma_c$ and unstable immediately above, moreover also the bifurcated branches are unstable. In our  model the bifurcated configuration does not develop oscillations, it has only one fold that may have a large amplitude. This is because the model is one-dimensional; in fact if the model would have been two-dimensional  the boundary conditions would have constrained the surface displacement, forcing it to oscillate and fold \cite{Healey}.
Within this line of thought we may infer that the lack of stable branches for $\sigma$ slightly larger than $\sigma_c$ is  a manifestation of the fact that not only creases form but also that the crease tip folds up and forms self-contact \cite{vanderHeijden2003,Jin2015,ciarletta2018,ciarletta2019}.

To show that there is a branch that bifurcates from the straight configuration for $\sigma=\sigma_c$, we evaluate the derivatives of $\beta$ at $\eta=0$. We will show that the bifurcation is not \textit{transcritical} (\textit{i.e.} $\beta'(0)=0$) but it is \textit{subcritical} (\textit{i.e.} $\beta''(0)<0$).
The computation of $\beta''(0)$ turns out to be quite involved since a linear operator needs to be inverted. In several standard problems this procedure is not necessary thanks to some symmetries enjoyed by the energy: symmetries that in the case under study are lacking.

Differentiating \eqref{fb} we find
\begin{equation}\label{bp}
\beta'(0)=-\frac 12 \frac{\partial^2 f/\partial \eta^2}{\partial^2 f/\partial \eta\partial\beta}(0,0).
\end{equation}
The denumerator has been computed in \eqref{fab}, while in  \ref{appf} it is shown that
\begin{equation}\label{faa}
\frac{\partial^2 f}{\partial\eta^2}(0,0)= d^3\Ec(\oo,\sigma_c)[\uu_c,\uu_c,\uu_c].
\end{equation}
To evaluate the right hand side of \eqref{faa}, we note that the  third derivatives of the density energy $w$
evaluated at $(\mu_0,0,0)$ are zero, apart from  the following:

\begin{equation}\label{d3w}
\begin{array}{ll}
\displaystyle\DD_{1}=\partial^3_{111}w(\mu_0,0,0)=-\frac{6}{\mu_0^5}\big(\varepsilon\big( 4-(4-3\alpha)\mu_0 \big)+\gamma\mu_0\big),
&
\displaystyle\DD_{2}=\partial^3_{221}w(\mu_0,0,0)=\frac{\varepsilon^2\gamma}{4},
\\
\displaystyle\DD_{3}=\partial^3_{223}w(\mu_0,0,0)=-\frac{\sigma\gamma\varepsilon^3\mu_0^3}{8},
&
\displaystyle\DD_{4}=\partial^3_{331}w(\mu_0,0,0)=\frac{\varepsilon^3\mu_0}{3}.
\end{array}
\end{equation}
We can therefore write the right hand side of \eqref{faa} as follows:
\begin{equation}\label{d3E}
\begin{aligned}
d^3\Ec(\oo,\sigma_c)[\uu_c,\uu_c,\vv]=&\int_0^1\DD_{1}(\xi_c)^2\xi+
\DD_{2}\big((\xi_c^{\prime})^2\xi+2\xi_c\xi_c^{\prime}\xi'\big)\\
&\qquad+
\DD_{3}\big((\xi_c^{\prime})^2\vartheta'+2\xi_c^{\prime}\vartheta_c^{\prime}\xi'\big)\\
&\qquad+
\DD_{4}\big((\vartheta_c^{ \prime})^2\xi+2\xi_c\vartheta_c^{\prime}\vartheta'\big)\,{\rm d}x,
\end{aligned}
\end{equation}
for $\vv=(\xi,\vartheta).$ In particular, in the computation of $d^3\Ec(\oo,\sigma_c)[\uu_c,\uu_c,\uu_c]$ we have to evaluate the integrals of the functions $(\xi_c)^3, (\xi_c^{\prime})^2\xi_c, (\xi_c^{\prime})^2\vartheta_c^{\prime}$, and $(\vartheta_c^{ \prime})^2\xi_c$, which are all odd functions respect to $x=1/2$, hence

\begin{equation}\label{b1}
\frac{\partial^2 f}{\partial\eta^2}(0,0)= d^3\Ec(\oo,\sigma_c)[\uu_c,\uu_c,\uu_c]=0.
\end{equation}
From \eqref{bp} we deduce that $\beta'(0)=0$, which means that the bifurcation is not \textit{transcritical}.

We now have to determine $\beta''(0)$. Differentiating three times \eqref{fb}, we deduce that the second derivative of $\beta$ is
\begin{equation}\label{bpp}
\beta''(0)=-\frac 13 \frac{\partial^3 f/\partial \eta^3}{\partial^2 f/\partial \eta\partial\beta}(0,0).
\end{equation}
Again, the denumerator has been computed in \eqref{fab}, while in  \ref{appf} it is shown that
$$\frac{\partial^3 f}{\partial\eta^3}(0,0)=d^4\Ec(\oo,\sigma_c)[\uu_c,\uu_c,\uu_c,\uu_c]
+3d^3\Ec(\oo,\sigma_c)[\uu_c,\uu_c,\uu_c+\tilde \uu_{\eta\eta}],
$$
where
$$
\tilde \uu_{\eta\eta}=(\xi_{\eta\eta},\vartheta_{\eta\eta})=\frac{\partial^2\tilde\uu}{\partial\eta\partial\beta}(0,0),
$$
and  $\tilde \uu$  defined in \eqref{vsol}. 
By taking into account \eqref{b1}, the above equation reduces to
\begin{equation}\label{faaa}
\frac{\partial^3 f}{\partial\eta^3}(0,0)=d^4\Ec(\oo,\sigma_c)[\uu_c,\uu_c,\uu_c,\uu_c]
+3d^3\Ec(\oo,\sigma_c)[\uu_c,\uu_c,\tilde \uu_{\eta\eta}].
\end{equation}

To compute the fourth variation of the energy, we note that  the  fourth derivatives of the density energy $w$
evaluated at $(\mu_0,0,0)$ are all zero, apart from:
\begin{equation}\label{d4w}
\begin{aligned}
&\BB_{1}=\partial^4_{1111}w(\mu_0,0,0)=\frac{24}{\mu_0^6}\big(\varepsilon(5-(4-3\alpha)\mu_0)+\gamma\mu_0\big),
\quad
\BB_{2}=\partial^4_{1133}w(\mu_0,0,0)=\frac{\varepsilon^3}{3},\\
&\BB_{3}=\partial^4_{2213}w(\mu_0,0,0)=-\frac{3\gamma\varepsilon^3\mu_0^2}{8}\,\sigma_c.
\end{aligned}
\end{equation}

Taking into account \eqref{Nuc} we find 
\begin{equation}\label{d4E}
\begin{aligned}
d^4\Ec(\oo,\sigma_c)[\uu_c,\uu_c,\uu_c,\uu_c]&=\int_0^1\BB_{1}(\xi_c)^4+
6\BB_{2}(\xi_{c})^2(\vartheta_c^{\prime})^2
+12\BB_{3}\xi_c(\xi_c^{\prime})^2\vartheta_c^{\prime}\,{\rm d}x\\
&=\frac 38\BB_{1}+
9\pi^2\K^2 \BB_{2}+12\pi^3\K\BB_{3}.
\end{aligned}
\end{equation}

In order to evaluate the second term appearing in \eqref{faaa}, we need to find $\tilde \uu_{\eta\eta}$. To this end we differentiate \eqref{eqb3} twice with respect to $\eta$ to find
\begin{equation}\label{vaa}
d^3\Ec(\oo,\sigma_c)[\uu_c,\uu_c,\tilde\vv]+d^2\Ec(\oo,\sigma_c)[\tilde \uu_{\eta\eta},\tilde \vv]=0\quad \forall\tilde \vv\in\NN^\perp.
\end{equation}
Problem \eqref{vaa} admits a unique solution, since the bi-linear form $d^2\Ec(\oo,\sigma_c)$ is positive definite on $\NN^\perp$.  

One of the major problems in evaluating \eqref{vaa} lies in the computation of $\tilde \uu_{\eta\eta}$, because it is necessary to invert a linear operator. In a number of usual applications, this computation is not necessary, as $d\Ec(\uu,\sigma)[\vv]$ is an odd function, \textit{i.e.}, $d\Ec(-\uu,\sigma)[\vv]=-d\Ec(-\uu,\sigma)[\vv]$. If this was the case, we would have 
$
d^3\Ec(\oo,\sigma_c)[\uu_c,\uu_c,\tilde \uu_{\eta\eta}]=0.
$
The present problem does not fall within this circumstance. In  \ref{appvaa} we carry out the cumbersome computation of  $\tilde \uu_{\eta\eta}$; in particular, we show that
\begin{equation}
\begin{aligned}
&\xi_{\eta\eta}=\Upsilon_1\cos(2\pi x)+\Upsilon_2\cos(4\pi x)+\Upsilon_3,\\
&\vartheta_{\eta\eta}=\Psi_1\sin(2\pi x)+\Psi_2 \sin(4\pi x),\\
\end{aligned}
\end{equation}
where the constants $\Upsilon_i$ and $\Psi_i$ are defined in \eqref{Upsi}.

With this, we are in position to evaluate 
\begin{equation}
d^3\Ec(\oo,\sigma_c)[\uu_c,\uu_c,\tilde \uu_{\eta\eta}]=-\frac{1}{4}\Gamma_1(\Upsilon_1-2\Upsilon_3)+\pi(\Gamma_3\Psi_2+\pi\Gamma_2(\Upsilon_2+2\Upsilon_3)),
\end{equation}
and finally, by using \eqref{fab}, \eqref{d4E}, we can determine $\beta''(0)$ from \eqref{bpp}:
\begin{equation}
\begin{aligned}\label{betas}
\beta''(0)=-\frac{1}{8\pi\gamma\varepsilon\kappa}\Big( & \Lambda_1-2\Gamma_1(\Upsilon_2-2\Upsilon_3)+8\pi^2\big(-3\Gamma_2\Upsilon_2+4\pi\kappa(\Lambda_3-2\Gamma_3\Upsilon_2) \\
& +2\Gamma_2\Upsilon_3 +\kappa^2(3\Lambda_2-\Gamma_4\Upsilon_2+2\Gamma_4\Upsilon_3)+4\pi\Gamma_3\Psi_2-4\Gamma_4\Psi_2 \big)\Big).
\end{aligned}
\end{equation}

\begin{figure}[h!]
	\centering
	\includegraphics[scale=.5]{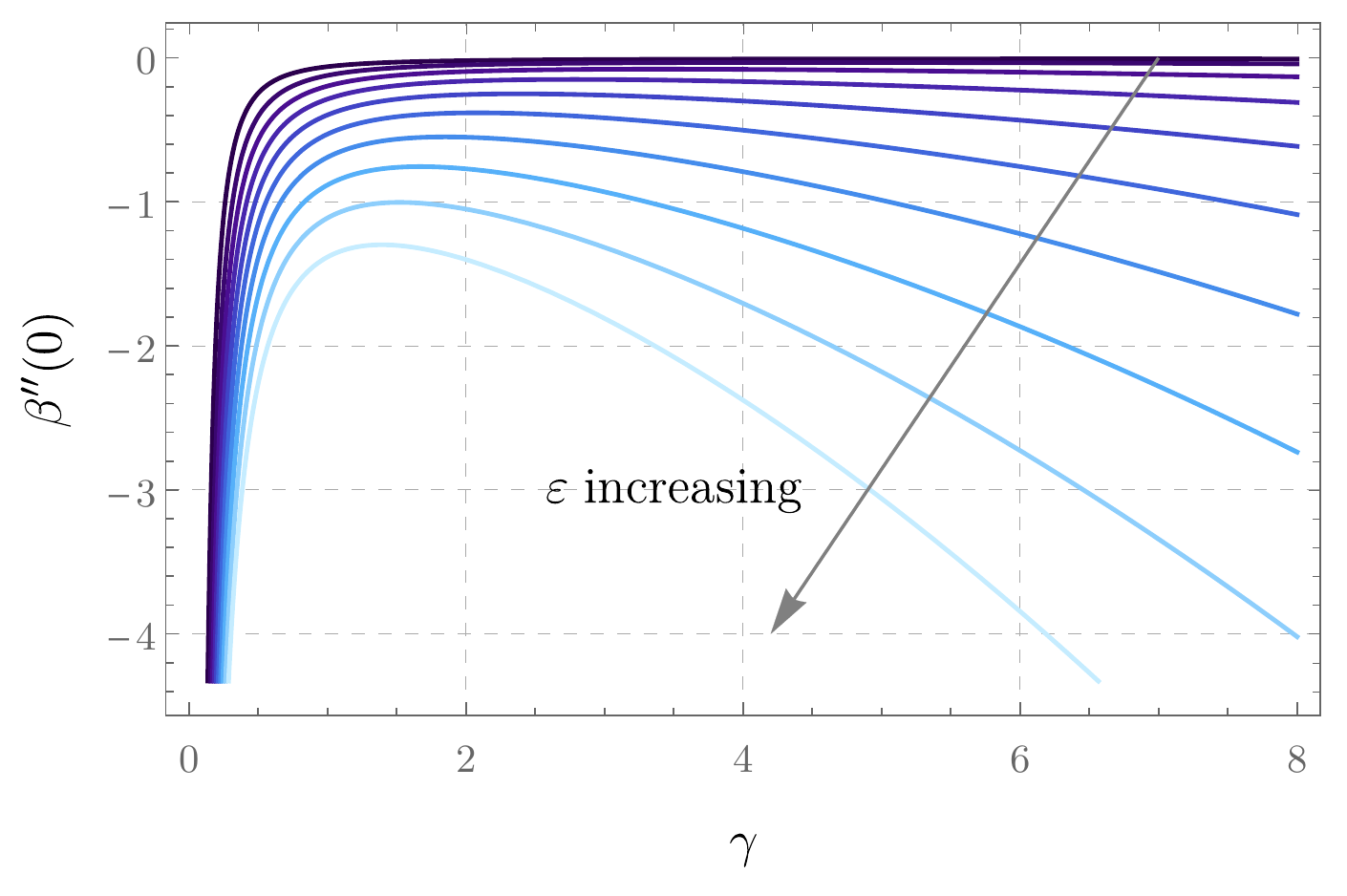}
	\caption{$\beta''(0)$, for given $\alpha$, in terms of $\gamma$ and for different values of $\varepsilon$}
	\label{fig:betas}
\end{figure}

%
In Fig. \eqref{fig:betas} we  report the function $\beta''(0)$, for given $\alpha$, in terms of $\gamma$ and for different values of $\varepsilon$. Since $\beta''(0)$ is always negative, we conclude that \textit{the bifurcation is subcritical}.


To gain further insight, we have computed the points of the solution branch using Keller's pseudo-arclength continuation \cite{keller1977} implemented in the software \texttt{AUTO-07p} \cite{doedel2007auto}.  In the numerical example, we fixed $\varepsilon=0.1$ and, based on Table \ref{tablee}, we chose  $\gamma=5$, so that the critical value $\sigma_c$ is approximately the intermediate value within the admissible range $[0,1/2]$. We further set  $\alpha=0.2$.

\begin{figure}[h!]
	\centering
	\includegraphics[width=0.7\linewidth]{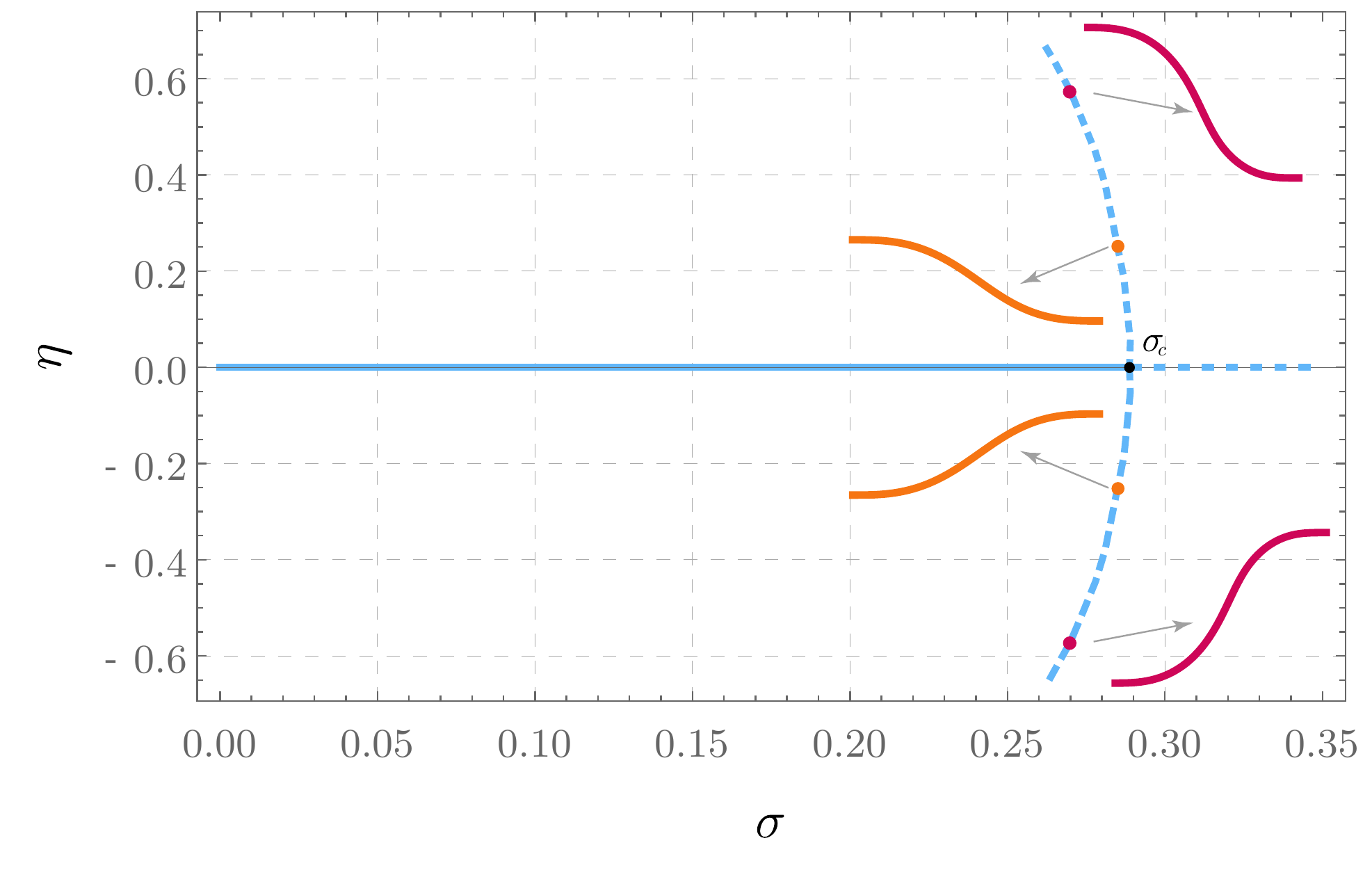}
	\caption{Symmetric branch emanating from the bifurcation point $(0,\sigma_c)$. The points marked on the upper side have coordinates $(0.21,0.58)$ and $(0.20,1.21)$. The corresponding deformed configurations are also shown.}
	\label{fig:biforcazione_GT}
\end{figure}

\begin{figure}[h!]
	\centering
	\includegraphics[scale=.32]{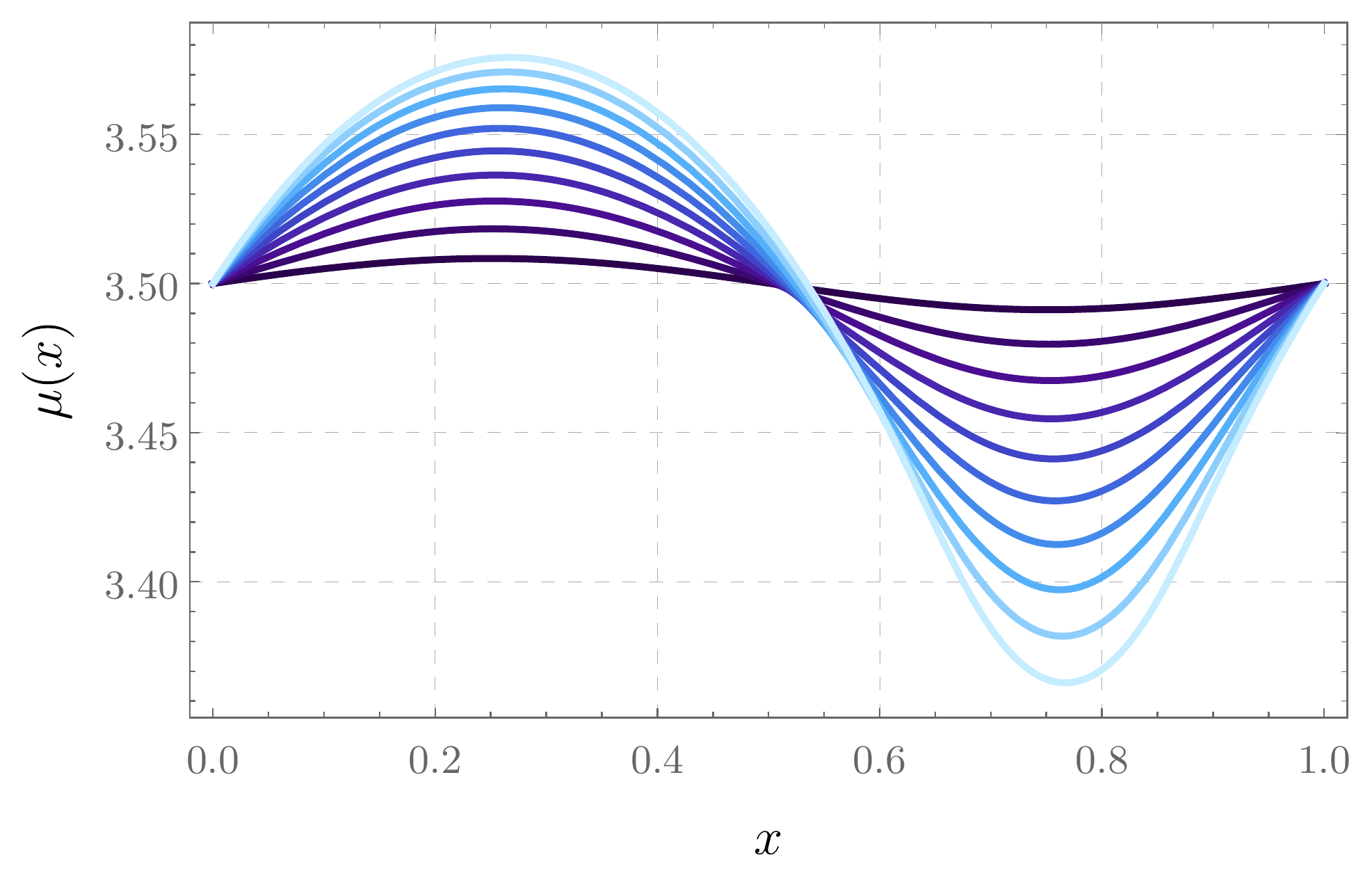}
	\includegraphics[scale=.32]{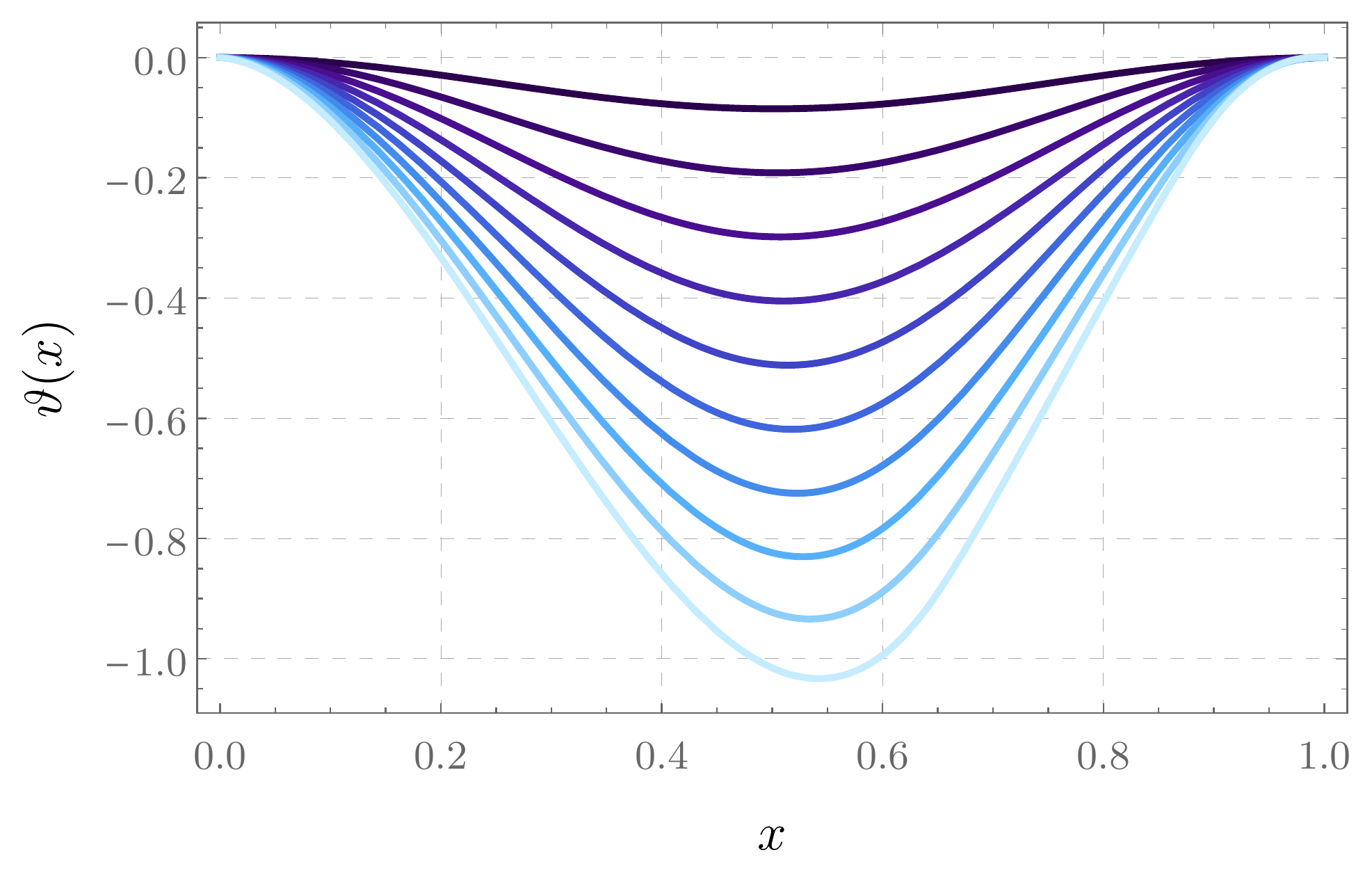}
	\includegraphics[scale=.32]{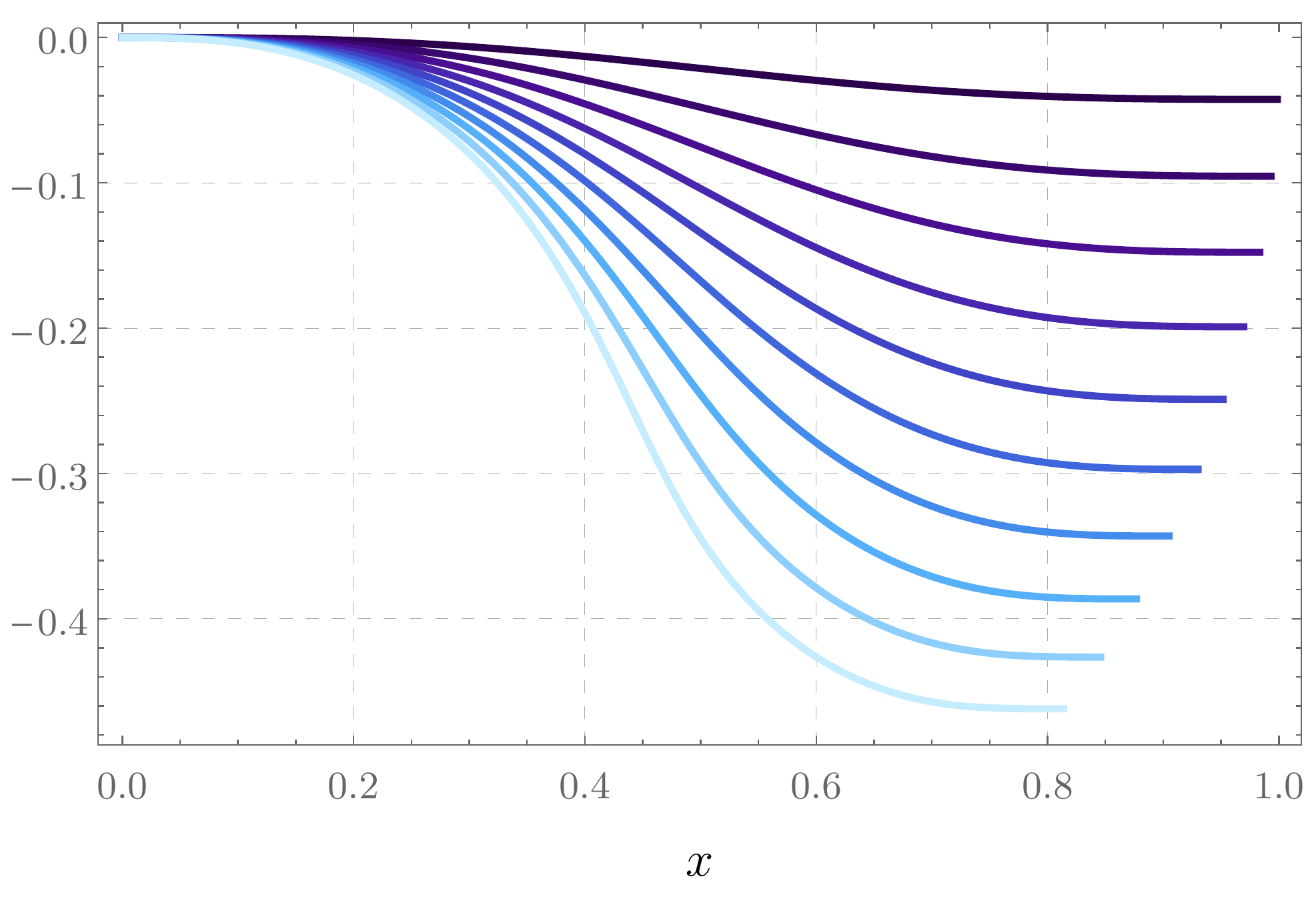}
	\caption{Numerical solutions of $\mu(x)$, $\vartheta(x)$ and shape of the midline for different  values of $\sigma$.}
	\label{fig:sols}
\end{figure}

We obtained the diagram shown in Fig. \ref{fig:biforcazione_GT} where the symmetric bifurcated paths stop at two endpoints where \texttt{AUTO} failed to converge. 
For completeness we also report some equilibrium configurations for different values of $\sigma$ (Fig. \ref{fig:sols}). Although such equilibria are not stable, and hence cannot be observed as stationary states, they may in principle represent snapshots along an  evolution path, provided that the model is endowed with some dynamics.

\section{Kras oncogene activation makes epithelial cells softer }\label{sec:kras-oncog-activ}
In this section we show that our theory predicts a distinctive mechanical behavior of pre-cancerous cells.

Our analysis is based on the data available in \cite{Messal}, where a technique for 3D imaging, named fast light-microscopic analysis of antibody-stained whole organ (FLASH), is set to perform Immunofluorescence measures on dissected pancreatic tissues of mice where \textit{Kras oncogene} activation is induced. 

\begin{figure}[h]
	\centering
	\includegraphics[scale=.45]{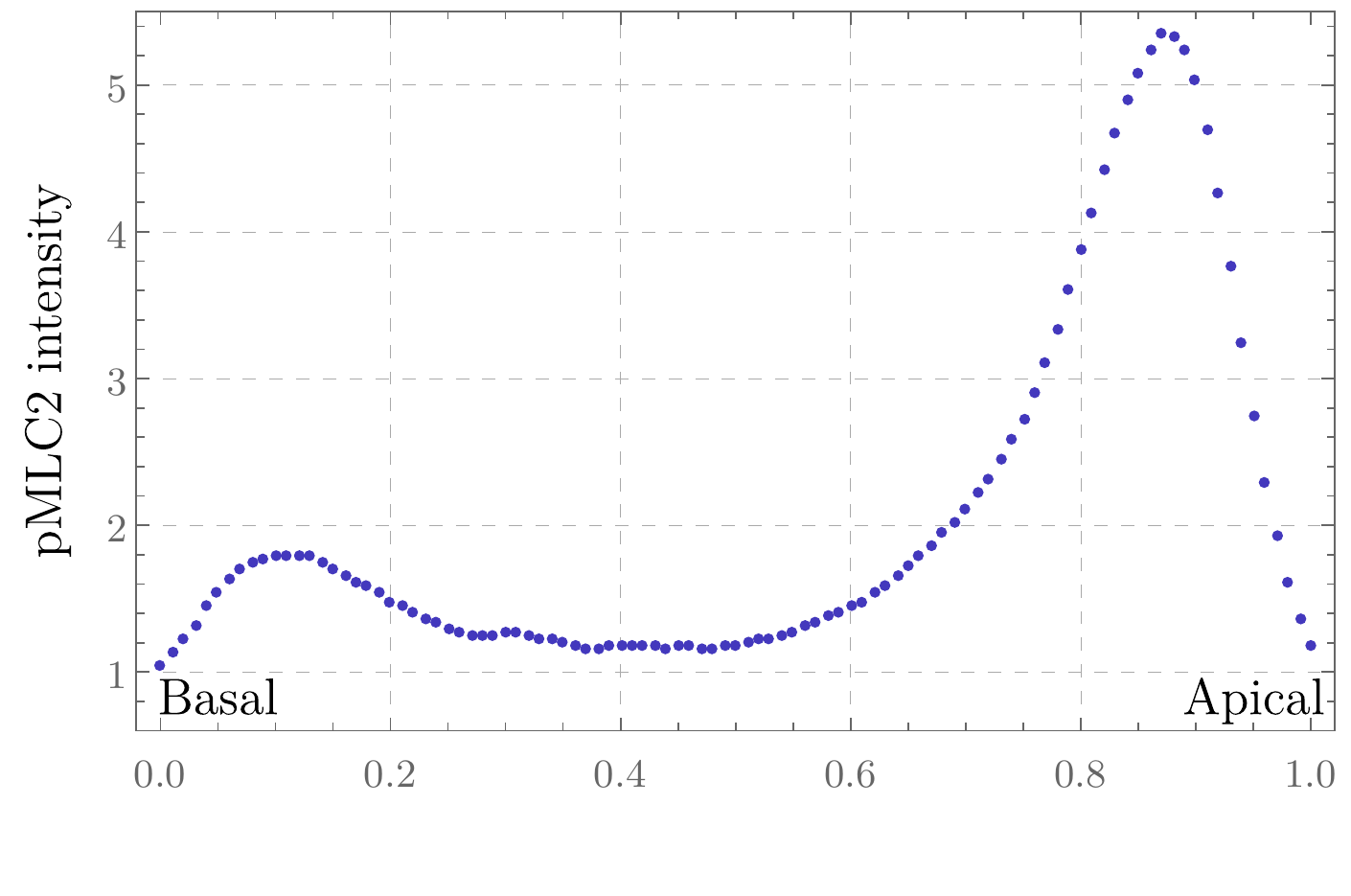}
	\includegraphics[scale=.45]{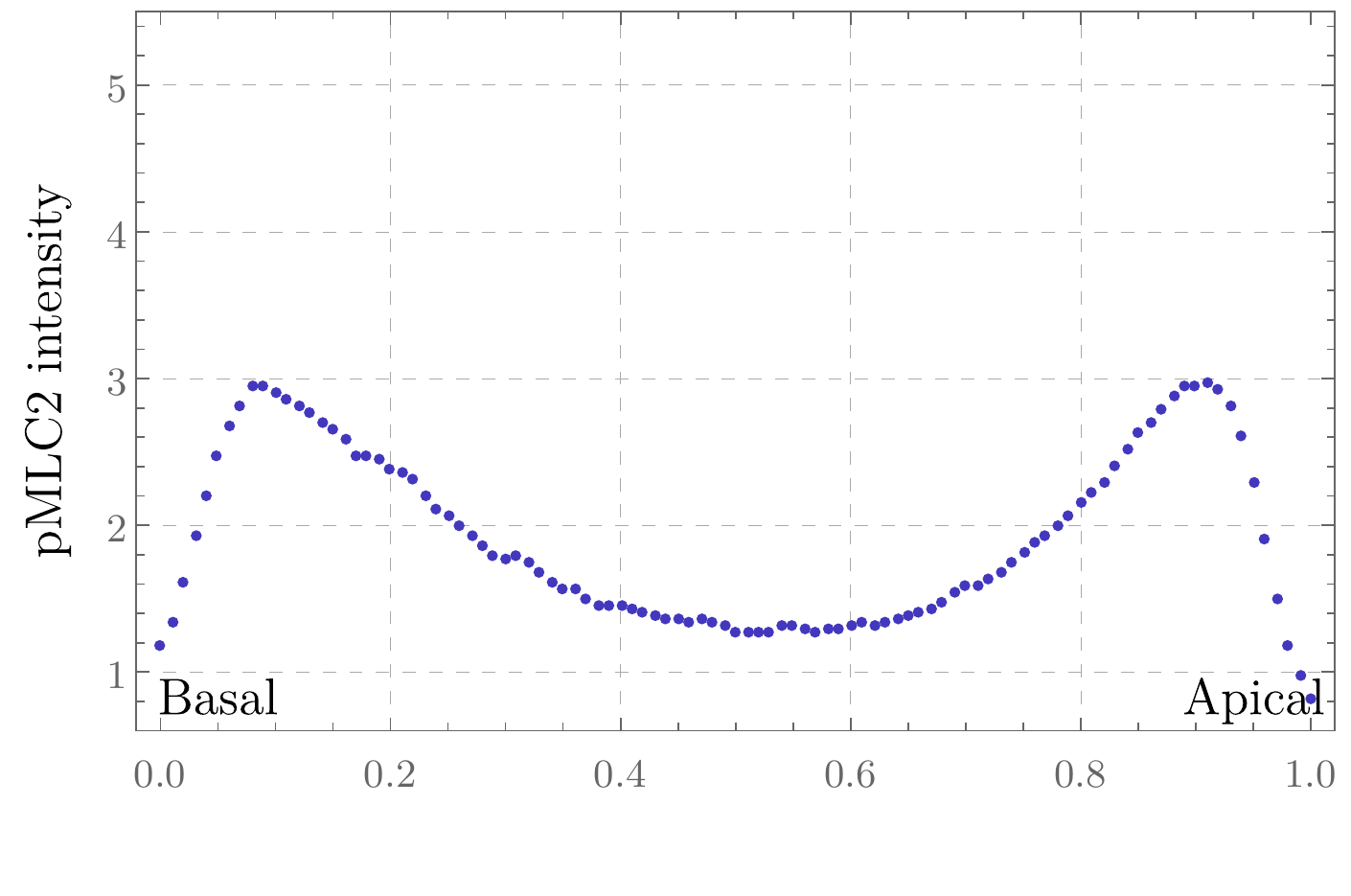}
	\caption{Concentration of Phospho-Myosin Light Chain 2  (pMLC2) in mice pancreatic epithelial tissue,  healthy (left),  transformed  (right). Data from \cite{Messal}. }
	\label{fig:messal_transformed_exo}
\end{figure}

The authors measure phosphorylated myosin light chain (pMLC2) and find that after oncogene activation ductal structures have the same pMLC2 distribution in the apical and basal regions, which is consistent with the observed distribution of F-actin. In Fig. \eqref{fig:messal_transformed_exo} we reported the data from \cite{Messal}, measured in approximately 70 healthy cells and 70 transformed cells from 7 mice.


The  reported pMLC2 intensities values are:
$$
I^h_a=5.3,\qquad I^t_a=2.9,\qquad I^h_b=1,8,\qquad I^t_b=2.9,
$$
where the superscripts  $h$ and $t$ denote healthy  (before transformation) and transformed cells, while the subscripts $a$ and $b$ indicate the apical and basal side, respectively.
As in  \cite{Messal}, we assume that apical and basal surface tensions are proportional to pMLC2 intensities:
\begin{equation}\label{pmlc2}
\sigma_a^h= 5.34 \varpi, \qquad \sigma_a^t= 2.96 \varpi \qquad \sigma_b^h= 1.80 \varpi,  \qquad \sigma_b^t= 2.96 \varpi,
\end{equation}
where $\varpi$ is a proportionality constant.
Recalling the definition of $\sigma$ \eqref{const}$_3$, we find
\begin{equation}
\sigma^h=\frac 12 \frac{\sigma_{a}^h-\sigma_b^h}{\sigma_{a}^h+\sigma_{b}^h}=0.25,
\quad
\sigma^t=\frac 12 \frac{\sigma_{a}^t-\sigma_b^t}{\sigma_{a}^t+\sigma_{b}^t}=0.
\end{equation}

From \href{https://www.nature.com/articles/s41586-019-0891-2/figures/2}{Fig. 2l} of  \cite{Messal} we deduce that the ratio between the thickness of transformed ducts, $\mu^t$, and a healthy ducts, $\mu^h$, is
$$
\chi=\frac{\mu^t}{\mu^h}=1.3.
$$ 
Assuming, that the $\mu$ values differ only slightly from $\mu_0$ (as we numerically observe) we deduce that
\begin{equation}\label{chi0}
\frac{\mu^t_0}{\mu^h_0}=\chi.
\end{equation}
For $\mu_0^3\gg 1$, and hence $\mu_0^4\gg \mu_0$, the quartic equation \eqref{mu0} simplifies to
$$
\mu_0^4  -\frac{\gamma}{2\varepsilon} \mu_0-\widehat\alpha\mu_0^3\simeq 0, 
$$
where 
$$
\widehat\alpha=\frac{4-3\alpha}{2}
$$
is always greater than $1/2$ and smaller than $2$.
Simplifying $\mu_0$ and adding the negligible term $3\mu_0(\widehat \alpha/3)^2-(\widehat \alpha/3)^3$ to the previous identity we find
$$
\mu_0^3 -\widehat\alpha\mu_0^2+3\mu_0(\frac{\widehat \alpha}3)^2-(\frac{\widehat \alpha}3)^3\simeq \frac{\gamma}{2\varepsilon}, 
$$
that is
$$
(\mu_0 -\frac{\widehat \alpha}3)^3\simeq \frac{\gamma}{2\varepsilon}. 
$$
Thus
\begin{equation}\label{approx_mu0}
\mu_0\simeq \frac{\widehat \alpha}3 + \big(\frac{\gamma}{2\varepsilon}\big)^\frac 13.
\end{equation}
Equation \eqref{approx_mu0}, which has been tested for several values of $\alpha, \gamma$, and $\varepsilon$, allows to write \eqref{chi0} as
$$
\chi=\frac{\frac{\widehat \alpha^t}3 + \big(\frac{\gamma^t}{2\varepsilon}\big)^\frac 13}{\frac{\widehat \alpha^h}3 + \big(\frac{\gamma^h}{2\varepsilon}\big)^\frac 13}=\frac{\widehat \alpha^t (2\varepsilon)^\frac 13 + 3(\gamma^t)^\frac 13}{\widehat \alpha^h (2\varepsilon)^\frac 13 + 3(\gamma^h)^\frac 13}\simeq \frac{(\gamma^t)^\frac 13}{ (\gamma^h)^\frac 13},
$$
since $\varepsilon$ is small and $1/2\le \widehat \alpha^t, \widehat \alpha^h \le 2$.
We therefore have found that
$$
\chi^3\simeq\frac{\gamma^t}{\gamma^h}.
$$
From the definition of $\gamma$ \eqref{const}$_4$ and the data in \eqref{pmlc2}, we  find
$$
\frac{5.92 \varpi}{\ell (\alpha_1^t+\alpha_2^t)}=\frac{\sigma_a^t+\sigma_b^t}{\ell (\alpha_1^t+\alpha_2^t)}\simeq\chi^3\frac{\sigma_a^h+\sigma_b^h}{\ell (\alpha_1^h+\alpha_2^h)}=1.3^3 \frac{7.14\varpi}{\ell (\alpha_1^h+\alpha_2^h)},
$$
which leads to
$$
\alpha_1^h+\alpha_2^h\simeq\frac{7.14 \ 1.3^3}{5.92}(\alpha_1^t+\alpha_2^t)=2.62 (\alpha_1^t+\alpha_2^t).
$$
Thus, our theory predicts that  in pancreatic epithelial ducts, shortly after oncogene activation, the \textit{transformed cells are softer than healthy ones}.

This finding agrees with \cite{Therville-2019}, where it is shown that microscopic tissural modifications during pretumoral stage, owed only to abnormal apico-basal differential tension, cause a softening of the pancreatic tissue, possibly leading to a change in the physiological folded morphology.

\appendix
\section{Properties of the reduced function $f(\eta,\beta)$}\label{appf}

%
%
In order to evaluate the derivative of the reduced function $f(\eta,\beta)$ with respect to $\eta$, we  note that, by \eqref{deff},
$$
\frac{\partial f}{\partial\eta}(\eta,\beta)=d^2\Ec(\eta\uu_c+\tilde \uu(\eta,\beta),\sigma_c +\beta)[\uu_c,\uu_c+\frac{\partial\tilde \uu}{\partial\eta}(\eta,\beta)],
$$
hence, in particular,
\begin{equation}\label{fa}
\frac{\partial f}{\partial\eta}(0,0)=d^2\Ec(0,\sigma_c)[\uu_c,\uu_c+\frac{\partial\tilde \uu}{\partial\eta}(0,0)]=0
\end{equation}
thanks to \eqref{uc}.

To continue, we need to note that 
\begin{equation}\label{dv}
\frac{\partial\tilde \uu}{\partial\beta}(0,\beta)=0 \quad\mbox{and}\quad
\frac{\partial\tilde \uu}{\partial\eta}(0,0)=0.
\end{equation}
The first follows directly from \eqref{eqb4}, while to show the second we differentiate \eqref{eqb3} with respect to $\eta$ to find
$$
0=\frac{\partial}{\partial\eta}d\Ec(\eta\uu_c+\tilde \uu(\eta,\beta),\sigma_c +\beta)[\tilde\vv]\rvert_{\eta=\beta=0}=d^2\Ec(0,\sigma_c)[\tilde\vv,\uu_c+\frac{\partial\tilde \uu}{\partial\eta}(0,0)],
$$
where we used  \eqref{eqb4}. Using \eqref{uc}, the above equation reduces to
$$
d^2\Ec(\oo,\sigma_c)[\tilde\vv,\frac{\partial\tilde \uu}{\partial\eta}(0,0)]=0,
$$
which implies the second of \eqref{dv}, since $\partial\tilde\uu/\partial\eta\in\NN^\perp$.

Differentiating \eqref{fa} with respect to $\beta$ we find
$$
\begin{aligned}
\frac{\partial^2 f}{\partial\eta\partial\beta}(\eta,\beta)=& \ d^3\Ec(\eta\uu_c+\tilde\uu(\eta,\beta),\sigma_c +\beta)[\uu_c,\uu_c+\frac{\partial\tilde\uu}{\partial\eta}(\eta,\beta),\frac{\partial\tilde\uu}{\partial\beta}(\eta,\beta)]\\
&+d^2\Ec(\eta\uu_c+\tilde\uu(\eta,\beta),\sigma_c +\beta)[\uu_c,\uu_c+\frac{\partial^2\tilde\uu}{\partial\eta\partial\beta}(\eta,\beta)]\\
&+d^2\Ec_\sigma(\eta\uu_c+\tilde\uu(\eta,\beta),\sigma_c +\beta)[\uu_c,\uu_c+\frac{\partial\tilde\uu}{\partial\eta}(\eta,\beta)].
\end{aligned}
$$
On taking into account \eqref{dv}, we find
$$
\begin{aligned}
\frac{\partial^2 f}{\partial\eta\partial\beta}(0,0)=& \ d^3\Ec(\oo,\sigma_c)[\uu_c,\uu_c0]
+d^2\Ec(\oo,\sigma_c)[\uu_c,\uu_c+\frac{\partial^2\tilde\uu}{\partial\eta\partial\beta}(0,0)]\\
&+d^2\Ec_\sigma(\oo,\sigma_c)[\uu_c,\uu_c].
\end{aligned}
$$
Eq. \eqref{uc} finally allows to conclude
$$
\frac{\partial^2 f}{\partial\eta\partial\beta}(0,0)=d^2\Ec_\sigma(\oo,\sigma_c)[\uu_c,\uu_c].
$$
With similar calculations we find
$$
\begin{aligned}
\frac{\partial^2 f}{\partial\eta^2}(0,0)=& \ d^3\Ec(\oo,\sigma_c)[\uu_c,\uu_c,\uu_c]
+d^2\Ec(\oo,\sigma_c)[\uu_c,\uu_c+\tilde{\uu}_{\eta\eta}]\\
= & \ d^3\Ec(\oo,\sigma_c)[\uu_c,\uu_c,\uu_c],\\
\frac{\partial^3 f}{\partial\eta^3}(0,0)=& \ d^4\Ec(\oo,\sigma_c)[\uu_c,\uu_c,\uu_c,\uu_c]
+3d^3\Ec(\oo,\sigma_c)[\uu_c,\uu_c,\uu_c+\tilde{\uu}_{\eta\eta}]\\
& \quad +d^2\Ec(\oo,\sigma_c)[\uu_c,\uu_c+\frac{\partial^3\tilde\uu}{\partial\eta^3}(0,0)]\\
= & \ d^4\Ec(\oo,\sigma_c)[\uu_c,\uu_c,\uu_c,\uu_c]
+3d^3\Ec(\oo,\sigma_c)[\uu_c,\uu_c,\uu_c+\tilde{\uu}_{\eta\eta}],
\end{aligned}
$$
where both equations have been simplified by using \eqref{uc}.

\section{Evaluation of $\tilde{\uu}_{\eta\eta}$}\label{appvaa}

For a generic function $\ww$ we define
$$
\langle \ww\rangle:=\frac 1{\int_0^1 \uu_c\cdot\uu_c\, {\rm d}x}\int_0^1 \uu_c\cdot\ww\, {\rm d}x,
$$
and
\begin{equation}\label{wt}
\tilde\ww:=\ww-\langle \ww\rangle\uu_c\in \NN^\perp.
\end{equation}
Clearly all functions in $\NN^\perp$ admit such a representation.

The problem that determines $\tilde\uu_{\eta\eta}$ is \eqref{vaa}:
\begin{equation}\label{vaa2}
d^3\Ec(\oo,\sigma_c)[\uu_c,\uu_c,\tilde\ww]+d^2\Ec(\oo,\sigma_c)[\tilde\uu_{\eta\eta},\tilde \ww]=0\quad \forall\tilde \ww\in\NN^\perp.
\end{equation}
With $\tilde\ww$ as in \eqref{wt} we have
$$
\begin{aligned}
d^3\Ec(\oo,\sigma_c)[\uu_c,\uu_c,\tilde\ww]& =d^3\Ec(\oo,\sigma_c)[\uu_c,\uu_c,\ww]-\langle \ww\rangle d^3\Ec(\oo,\sigma_c)[\uu_c,\uu_c,\uu_c]\\
& =d^3\Ec(\oo,\sigma_c)[\uu_c,\uu_c,\ww]
\end{aligned}
$$
by \eqref{b1}, and
$$
\begin{aligned}
d^2\Ec(\oo,\sigma_c)[\tilde\uu_{\eta\eta},\tilde \ww] & =d^2\Ec(\oo,\sigma_c)[\tilde\uu_{\eta\eta},\ww]-\langle \ww\rangle d^2\Ec(\oo,\sigma_c)[\tilde\uu_{\eta\eta},\uu_c]\\
& =d^2\Ec(\oo,\sigma_c)[\tilde\uu_{\eta\eta},\ww]
\end{aligned}
$$
by \eqref{uc}.
Thus, problem \eqref{vaa2} leads to
\begin{equation}\label{vaa3}
d^3\Ec(\oo,\sigma_c)[\uu_c,\uu_c,\ww]+d^2\Ec(\oo,\sigma_c)[\tilde\uu_{\eta\eta},\ww]=0\quad \forall \ww.
\end{equation}
This latter problem, defined in the whole space and not just in $\NN^\perp$, does not have a unique solution, in contrast with problem \eqref{vaa2}. Projecting on $\NN^\perp$, with \eqref{wt}, any solution of \eqref{vaa3} we find the unique solution of problem \eqref{vaa2}.

Integrating by parts \eqref{d3E} we find
$$
\begin{aligned}
d^3\Ec(\oo,\sigma_c)[\uu_c,\uu_c,\vv]=&\int_0^1
[
\DD_{1}(\xi_c)^2+\DD_{2}(\xi_c^{\prime})^2-
2\DD_{2}(\xi_c\xi_c^{\prime})'-2\DD_{3}(\xi_c^{\prime}\vartheta_c^{\prime})'\\
&\qquad+\DD_{4}(\vartheta_c^{ \prime})^2]\xi
-[\DD_{3}((\xi_c^{\prime})^2)'+2
\DD_{4}(\xi_c\vartheta_c^{\prime})']\vartheta\,{\rm d}x.
\end{aligned}
$$
On taking into account \eqref{Nuc}, we conclude 
$$
\begin{aligned}
d^3\Ec(\oo,\sigma_c)[\uu_c,\uu_c,\vv]=&\int_0^1
[\Xi_1+\Xi_2\sin^2(2\pi x)]\xi
+\Xi_3\sin(4\pi x)\vartheta\,{\rm d}x,
\end{aligned}
$$
where we have set
\begin{equation}
\begin{aligned}
&\Xi_1:=-4\pi^2(\DD_2+4\K\pi\DD_3),\quad
\Xi_2:=
\DD_{1}+12\pi^2\DD_{2}+32\K\pi^3\DD_{3}+4\K^2\pi^2\DD_{4},\\
&\Xi_3:=8\pi^2(\pi\DD_{3}-\K \DD_{4}).
\end{aligned}
\end{equation}

Similarly, from \eqref{d2E} we find
\begin{equation}\label{d2Eb}
d^{2} \mathcal{E}\left(\oo, \sigma_{c}\right)[\tilde\uu_{\eta\eta}, \overline{\vv}]=
\int_{0}^{1} 
[\CC_{11} {\xi}_{\eta\eta}+\sigma_{c} \CC_{13}{\vartheta}^{\prime}_{\eta\eta}-\CC_{22} {\xi}^{\prime\prime}_{\eta\eta} ]{\xi}
-[\sigma_{c} \CC_{13}{\xi}_{\eta\eta}^\prime+\CC_{33} {\vartheta}^{\prime\prime}_{\eta\eta}]
{\vartheta}\,d x
\end{equation}
and hence problem \eqref{vaa3} writes also as

\begin{equation}\label{d2d3}
\begin{cases}
\CC_{11} {\xi}_{\eta\eta}+\sigma_{c} \CC_{13}{\vartheta}^{\prime}_{\eta\eta}-\CC_{22} {\xi}^{\prime\prime}_{\eta\eta} =-\Xi_1-\Xi_2\sin^2(2\pi x),\\
\sigma_{c} \CC_{13}{\xi}_{\eta\eta}^\prime+\CC_{33} {\vartheta}^{\prime\prime}_{\eta\eta}=\Xi_3\sin(4\pi x).
\end{cases}
\end{equation}

On solving \eqref{d2d3}$_2$ for $\xi'_{\eta\eta}$ and inserting the result in \eqref{d2d3}$_1$, we get ${\xi}_{\eta\eta}$ in terms of  ${\vartheta}_{\eta\eta}$:
\begin{equation}\label{xaa}
{\xi}_{\eta\eta}=-\frac{1}{\CC_{11}}\left(\sigma_{c} \CC_{13} \vartheta^{ \prime}_{\eta\eta}+\frac{\CC_{22} \CC_{33}}{\sigma_{c} \CC_{13}} \vartheta^{ \prime \prime\prime}_{\eta\eta}+\Xi_1+\Xi_2\sin^2(2\pi x)-4\frac{\CC_{22}\pi}{\CC_{13}\sigma_c}\Xi_3\cos(4\pi x)\right).
\end{equation}
On inserting 	\eqref{xaa} in \eqref{d2d3}$_2$, we find the following problem for ${\vartheta}_{\eta\eta}$:
\begin{equation}\label{systaa}
\begin{cases}
\vartheta^{ \prime \prime \prime \prime}_{\eta\eta}+4\pi^2\vartheta^{ \prime \prime}_{\eta\eta}=\dfrac{ \CC_{11}\Xi_3+2\pi(8\pi\CC_{22}\Xi_3+\sigma_c\CC_{13}\Xi_2)}{\CC_{22}\CC_{33}}\sin(4\pi x) \quad \text { in }(0,1), \\
\vartheta_{\eta\eta}=0 \qquad \hfill \text { at } x=0,1, \\
\sigma_{c} \CC_{13} \vartheta^{ \prime}_{\eta\eta}+\dfrac{\CC_{22} \CC_{33}}{\sigma_{c} \CC_{13}} \vartheta^{ \prime \prime\prime}_{\eta\eta}+\Xi_1+\Xi_2\sin^2(2\pi x)-4\dfrac{\CC_{22}\pi}{\CC_{13}\sigma_c}\Xi_3\cos(4\pi x)=0 \hfill \text { at } x=0,1,
\end{cases}
\end{equation}
where we have made use of the relation
$$
4\pi^{2}=\frac{\sigma_c^{2} \CC_{13}^{2}-\CC_{11} \CC_{33}}{\CC_{22} \CC_{33}}.
$$

A particular solution of \eqref{systaa} is 
$$
\vartheta_{\eta\eta}^p(x)=k \sin(4\pi x),
$$
where 
$$
k:=-\frac{\CC_{11}\Xi_3+2\pi (8\pi\CC_{22}\Xi_3+\sigma_c\CC_{13}\Xi_2)}{192 \pi^4\CC_{22}\CC_{33}}.
$$
The solution of the homogeneous equation is
\begin{equation}
\vartheta^h_{\eta\eta}(x)=b_{1} \cos (2\pi x)+b_{2} \sin (2\pi x)+b_{3} x+b_{4},
\end{equation}
and then the general solution is 
\begin{equation}\label{thaa}
\vartheta_{\eta\eta}(x)=b_{1} \cos (2\pi x)+b_{2} \sin (2\pi x)+b_{3} x+b_{4}+k \sin(4\pi x),
\end{equation}
where the constants $b_1,\ldots,b_4$ have to be determined by enforcing the boundary conditions \eqref{systaa}$_{2,3}$.

From \eqref{xaa}, we get
\begin{equation}\label{xaa1}
{\xi}_{\eta\eta}(x)=b_2\Phi_1 \cos(2\pi x)+\Phi_2 \cos(4\pi x) + b_1\Phi_3\sin(2\pi x)+b_3\Phi_4+\Phi_5 , 
\end{equation}
where
\begin{equation}
\begin{aligned}
&\Phi_1=-\frac{1}{\CC_{11}}\left(  2\pi\sigma_c\CC_{13}-\frac{8\pi^3\CC_{22}\CC_{33}}{\CC_{13}\sigma_c}\right),\\
&\Phi_2=\frac{-1}{48\pi^3\sigma_c\CC_{11}\CC_{13}\CC_{22}\CC_{33}}\Big(  2\pi(8\pi\CC_{22}\Xi_3+\CC_{13}\Xi_2\sigma_c) (4\pi^2\CC_{22}\CC_{33}-\CC_{13}^2\sigma_c^2)+\\
&\hspace{1cm}\CC_{11}\Xi_3(16\pi^2\CC_{22}\CC_{33}-\CC_{13}^2\sigma_c^2) \Big),\\
&\Phi_3=-\frac{2}{\CC_{11}\CC_{33}\sigma_c}\Big(4\pi^3\CC_{22}\CC_{33}-\pi\CC_{13}^2\sigma_c^2\Big),   \\
&\Phi_4=-\frac{\CC_{13}}{\CC_{11}}\sigma_c,\\
&\Phi_5=-\frac{1}{2\CC_{11}}(2\Xi_1+\Xi_2).
\end{aligned}
\end{equation}

On using \eqref{thaa} and \eqref{xaa1}, the  boundary conditions \eqref{systaa}$_{2,3}$ read:
\begin{equation}\label{sys}
\begin{cases}
b_1+b_4=0\\
b_1 + b_3 + b_4=0\\
b_2\Phi_1 + b_3\Phi_4=-\Phi_2-\Phi_5 ,
\end{cases}
\end{equation}
since the two conditions \eqref{systaa}$_{3}$ coalesce. The system has $\infty^1$ solutions, parameterized by $b_4=-b_1$.  We  need to project the solution on $\NN^\perp$, \textit{i.e.}, we have to request the orthogonality condition
\begin{equation}
0=\int_{0}^1\xi_{\eta\eta}(x)\xi_{c}(x)+\vartheta_{\eta\eta}(x)\vartheta_{c}(x)\,{\rm d}x=\frac{1}{2}\big( \kappa(-b_1+b_3+2b_4)+b_1\Phi_3 \big),
\end{equation}
which determines the unique solution of \eqref{sys}, and then allows to conclude that
%
\begin{equation}\label{Upsi}
\begin{aligned}
&\vartheta_{\eta\eta}=\Psi_1\sin(2\pi x)+\Psi_2 \sin(4\pi x),\\
&\xi_{\eta\eta}=\Upsilon_1\cos(2\pi x)+\Upsilon_2\cos(4\pi x)+\Upsilon_3,
\end{aligned}
\end{equation}
with 
\begin{equation}
\begin{aligned}
\Psi_1=&\frac{1}{96\pi^4}\left( \frac{\CC_{11}+16\pi^2\CC_{22}}{\CC_{22}\CC_{33}} \Xi_3+  \frac{2\pi\CC_{13}\sigma_c}{\CC_{22}\CC_{33}}\Xi_2+\frac{12\pi^2(\CC_{11}\Xi_3+2\CC_{13}\pi\sigma_c(2\Xi_1+\Xi_2))}{4\pi^2\CC_{22}\CC_{33}-\CC_{13}^2\sigma_c^2}\right),\\
\Psi_2=&-\frac{1}{192\pi^4\CC_{22}\CC_{33}}\Big( \CC_{11}\Xi_3+2\pi(8\pi\CC_{22}\Xi_3+\CC_{13}\Xi_2\sigma_c) \Big),\\
\Upsilon_1=&\frac{1}{48\pi^3\sigma_c\CC_{11}\CC_{13}\CC_{22}\CC_{33}} \Big(  16\CC_{22}\CC_{33}\pi^2(\CC_{11}+4\CC_{22}\pi^2)\Xi_3+16\pi^3\sigma_{c}\CC_{13}\CC_{22}\CC_{33}(2\Xi_3+2\Xi_2)\\ 
&-\CC_{13}^2\sigma_c^2(\CC_{11}+16\pi^2\CC_{22})\Xi_3-2\pi\sigma_c^3\CC_{13}^2\Xi_2   \Big),  \\
\Upsilon_2=&\frac{1}{48\pi^3\sigma_c\CC_{11}\CC_{13}\CC_{22}\CC_{33}} \Big(  \CC_1\Xi_3(\CC_{13}^2\sigma_c^2-16\pi^2\CC_{22}\CC_{33}),\\ &-2\pi(8\pi\CC_{22}\Xi_3+\sigma_c\CC_{13}\Xi_2)(4\pi^2\CC_{22}\CC_{33}-\CC_{13}^2\sigma_c^2)\Big),\\
\Upsilon_3=&\frac{1}{2\CC_{11}}(2\Xi_1+\Xi_2).
\end{aligned}
\end{equation}


\noindent We can then compute 
\begin{equation}
d^3\Ec(\oo,\sigma_c)[\uu_c,\uu_c,\tilde\uu_{\eta\eta}]=-\frac{1}{4}\Gamma_1(\Upsilon_1-2\Upsilon_3)+\pi(\Gamma_3\Psi_2+\pi\Gamma_2(\Upsilon_2+2\Upsilon_3)).
\end{equation}
Eqs. \eqref{fab}, \eqref{bpp},  \eqref{d4E}, finally allow to determine $\beta''(0)$ as in \eqref{betas}.

%
%

\begin{thebibliography}{10}
	\expandafter\ifx\csname url\endcsname\relax
	\def\url#1{\texttt{#1}}\fi
	\expandafter\ifx\csname urlprefix\endcsname\relax\def\urlprefix{URL }\fi
	\expandafter\ifx\csname href\endcsname\relax
	\def\href#1#2{#2} \def\path#1{#1}\fi
	
	\bibitem{Weinberg-2013}
	R.~Weinberg, The Biology of Cancer, W.W. Norton \& Company, 20013.
	
	\bibitem{Therville-2019}
	N.~Therville, S.~Arcucci, A.~Vertut, F.~Ramos-Delgado, D.~Da~Mota, M.~Dufresne,
	C.~Basset, J.~Guillermet-Guibert, Experimental pancreatic cancer develops in
	soft pancreas: novel leads for an individualized diagnosis by ultrafast
	elasticity imaging, Theranostics 9 (2019) 6369--6379.
	
	\bibitem{lewis_mechanics_1947}
	W.~H. Lewis, Mechanics of invagination, The Anatomical Record 97 (1947)
	139--156.
	
	\bibitem{Honda-1980}
	H.~Honda, G.~Eguchi, How much does the cell boundary contract in a monolayered
	cell sheet?, Journal of Theoretical Biology 84~(3) (1980) 575--588.
	
	\bibitem{Nagai-Honda-2009}
	T.~Nagai, H.~Honda, A dynamic cell model for the formation of epithelial
	tissues, Philosophical Magazine B 81~(7) (2001) 699--719.
	
	\bibitem{Bergmann-2018}
	A.~Nestor-Bergmann, G.~Goddard, S.~Woolner, O.~Jensen, Relating cell shape and
	mechanical stress in a spatially disordered epithelium using a vertex-based
	model, Math Med Biol 35~(Issue Supplement-1) (2018) 1--27.
	
	\bibitem{latorre_active_2018}
	E.~Latorre, S.~Kale, L.~Casares, M.~Gomez-Gonzalez, M.~Uroz, L.~Valon, R.~V.
	Nair, E.~Garreta, N.~Montserrat, A.~del Campo, B.~Ladoux, M.~Arroyo,
	X.~Trepat, Active superelasticity in three-dimensional epithelia of
	controlled shape, Nature 563~(7730) (2018) 203--208.
	
	\bibitem{misra_shape_2016}
	M.~Misra, B.~Audoly, I.~Kevrekidis, S.~Shvartsman, Shape transformations of
	epithelial shells, Biophysical Journal 110~(7) (2016) 1670--1678.
	
	\bibitem{misra_complex_2017}
	M.~Misra, B.~Audoly, S.~Y. Shvartsman, Complex structures from patterned cell
	sheets, Phil. Trans. R. Soc. B 372~(1720).
	
	\bibitem{Murisic-2015}
	N.~Murisic, V.~Hakim, I.~G. Kevrekidis, S.~Y. Shvartsman, B.~Audoly, From
	discrete to continuum models of three-dimensional deformations in epithelial
	sheets, Biophysical Journal 109~(1) (2015) 154--163.
	
	\bibitem{Bielmeier-2016}
	C.~Bielmeier, S.~Alt, V.~Weichselberger, M.~L. Fortezza, H.~Harz,
	F.~J\''ulicher, G.~Salbreux, A.-K. Classen, Interface contractility between
	differently fated cells drives cell elimination and cyst formation, Current
	Biology 26~(5) (2016) 563--574.
	
	
	\bibitem{preziosi}
	P. Ciarletta, D. Ambrosi, G. A. Maugin, L. Preziosi ,	Mechano-transduction in tumour growth modelling, The European Physical Journal E, 36: 23 (2013).
	
	\bibitem{fraldi}
	A.R. Carotenuto, A. Cutolo, S. Palumbo, M. Fraldi, Lyapunov stability of competitive cells dynamics in tumor mechanobiology. Acta Mechanica Sinica 37, 244-263 (2021).
	
	
	
	\bibitem{pugno}
	M. Fraldi, A. Cugno, L. Deseri, K. Dayal, N.M. Pugno, A frequency-based hypothesis for mechanically targeting and selectively attacking cancer cells. J.  R.  Soc.Interface12: 20150656 (2015).
	
	\bibitem{Drasdo-2000}
	D.~Drasdo, Buckling instabilities of one-layered growing tissues, Physical
	Review Letters 84 (2000) 4244--4247.
	
	\bibitem{Shraiman-2005}
	B.~I. Shraiman, Mechanical feedback as a possible regulator of tissue growth,
	Proceedings of the National Academy of Sciences 102~(9) (2005) 3318--3323.
	
	\bibitem{Hohlfeld-2011}
	E.~Hohlfeld, L.~Mahadevan, Unfolding the sulcus, Physical Review Letters 106
	(2011) 105702.
	
	\bibitem{Li-2012}
	B.~Li, Y.-P. Cao, X.-Q. Feng, H.~Gao, Mechanics of morphological instabilities
	and surface wrinkling in soft materials: a review, Soft Matter 8 (2012)
	5728--5745.
	
	\bibitem{BenAmar-2013}
	M.~Ben~Amar, F.~Jia, Anisotropic growth shapes intestinal tissues during
	embryogenesis, Proceedings of the National Academy of Sciences 110~(26)
	(2013) 10525--10530.
	
	\bibitem{Balbi-2015}
	V.~Balbi, E.~Kuhl, P.~Ciarletta, Morphoelastic control of gastro-intestinal
	organogenesis: {Theoretical} predictions and numerical insights, Journal of
	the Mechanics and Physics of Solids 78 (2015) 493--510.
	
	\bibitem{Salbreux-2017}
	G.~Salbreux, F.~J\"ulicher, Mechanics of active surfaces, Physical Review E 96
	(2017) 032404.
	
	\bibitem{Destrade_2020}
	V.~Balbi, M.~Destrade, A.~Goriely, Mechanics of human brain organoids, Phys.
	Rev. E 101 (2020) 022403.
	
	\bibitem{Hannezo-2014}
	E.~Hannezo, J.~Prost, J.~Joanny, Theory of epithelial sheet morphology in three
	dimensions, Proceedings of the National Academy of Sciences U. S. A. 111
	(2017) 27--32.
	
	\bibitem{Krajnc-2013}
	M.~Krajnc, N.~\v{S}torgel, A.~Ho\v{c}evar~Brezav\v{s}\v{e}k, P.~Ziherl, A
	tension-based model of flat and corrugated simple epithelia, Soft Matter 9.
	
	\bibitem{Papastavrou-2013}
	A.~Papastavrou, P.~Steinmann, E.~Kuhl, On the mechanics of continua with
	boundary energies and growing surfaces, Journal of the Mechanics and Physics
	of Solids 61~(6) (2013) 1446--1463.
	
	\bibitem{Julicher-2018}
	L.~Sui, S.~Alt, M.~e.~a. Weigert, Differential lateral and basal tension drive
	folding of drosophila wing discs through two distinct mechanisms, Nature
	Communications 9.
	
	\bibitem{Krajnc-2015}
	M.~Krajnc, P.~Ziherl, Theory of epithelial elasticity, Physical Review E 92
	(2015) 052713.
	
	\bibitem{Krajnc-2016}
	N.~\v{S}torgel, M.~Krajnc, P.~Mrak, J.~\v{S}trus, P.~Ziherl, Quantitative
	morphology of epithelial folds, Biophysical Journal 110 (2016) 269--77.
	
	\bibitem{Farhadifar-2007}
	R.~Farhadifar, J.-C. R\"oper, B.~Aigouy, S.~Eaton, F.~J\"ulicher, The influence
	of cell mechanics, cell-cell interactions, and proliferation on epithelial
	packing, Current Biology 17~(24) (2007) 2095--2104.
	
	\bibitem{Krajnc-2020}
	V.~Fiore, M.~Krajnc, F.~e.~a. Quiroz, Mechanics of a multilayer epithelium
	instruct tumour architecture and function, Nature 585 (2020) 433--439.
	
	\bibitem{Messal}
	H.~A. Messal, S.~Alt, R.~e.~a. Ferreira, Tissue curvature and apicobasal
	mechanical tension imbalance instruct cancer morphogenesis, Nature 566 (2019)
	126--130.
	
	\bibitem{haas2019}
	P.~A. Haas, R.~E. Goldstein, Nonlinear and nonlocal elasticity in
	coarse-grained differential-tension models of epithelia, Physical Review E 99
	(2019) 022411.
	
	\bibitem{antman}
	S.~Antman, The theory of rods. in: Truesdell \uppercase{C}. (eds) linear
	theories of elasticity and thermoelasticity.
	
	\bibitem{antman-2005}
	S.~Antman, Nonlinear Problems of Elasticity, Springer-Verlag, New York, 2005.
	
	\bibitem{ppg1982}
	P.~Podio-Guidugli, Flexural instabilities of elastic rods, Journal of
	Elasticity 12~(1) (1982) 3--17.
	
	\bibitem{coman_2017}
	C.D.~Coman, Bifurcation instabilities in finite bending of circular cylindrical shells, International Journal of Engineering Science, 119 (2017) 249--264.
	
	
	
	\bibitem{mora2013}
	S.~Mora, C.~Maurini, T.~Phou, J.-M. Fromental, B.~Audoly, Y.~Pomeau, Solid
	drops: Large capillary deformations of immersed elastic rods, Physical Review
	Letters 111 (2013) 114301.
	
	\bibitem{bico2018}
	J.~Bico, V.~Reyssat, B.~Roman, Elastocapillarity: When surface tension deforms
	elastic solids, Annual Review of Fluid Mechanics 50~(1) (2018) 629--659.
	
	\bibitem{Hohlfeld2012}
	E.~Hohlfeld, L.~Mahadevan, Scale and nature of sulcification patterns, Physical
	Review Letters 109 (2012) 025701.
	
	\bibitem{Chen2014}
	D.~Chen, L.~Jin, Z.~Suo, R.~C. Hayward, Controlled formation and disappearance
	of creases, Materials Horizons 1 (2014) 207--213.
	
	\bibitem{Chen2012}
	D.~Chen, S.~Cai, Z.~Suo, R.~C. Hayward, Surface energy as a barrier to creasing
	of elastomer films: An elastic analogy to classical nucleation, Physical
	Review Letters 109 (2012) 038001.
	
	\bibitem{Healey}
	T.~Healey, Q.~Li, R.~Cheng, {Wrinkling Behavior of Highly Stretched Rectangular
		Elastic Films via Parametric Global Bifurcation}, Journal of Nonlinear
	Science 23 (2013) 777--805.
	
	\bibitem{vanderHeijden2003}
	G.~van~der Heijden, S.~Neukirch, V.~Goss, J.~Thompson, Instability and
	self-contact phenomena in the writhing of clamped rods, International Journal
	of Mechanical Sciences 45~(1) (2003) 161--196.
	
	\bibitem{Jin2015}
	L.~Jin, A.~Auguste, R.~C. Hayward, Z.~Suo, {Bifurcation Diagrams for the
		Formation of Wrinkles or Creases in Soft Bilayers}, Journal of Applied
	Mechanics 82~(6).
	
	\bibitem{ciarletta2018}
	P.~Ciarletta, Matched asymptotic solution for crease nucleation in soft solids,
	Nature Communications 9~(496).
	
	\bibitem{ciarletta2019}
	P.~Ciarletta, L.~Truskinovsky, Soft nucleation of an elastic crease, Phys. Rev.
	Lett. 122 (2019) 248001.
	
	\bibitem{keller1977}
	H.~Keller, Numerical solution of bifurcation and nonlinear eigenvalue problems,
	In: Rabinowitz, P., Ed., Applications of Bifurcation Theory, Academic Press,
	New York  359--384.
	
	\bibitem{doedel2007auto}
	E.~J. Doedel, A.~R. Champneys, F.~Dercole, T.~F. Fairgrieve, Y.~A. Kuznetsov,
	B.~Oldeman, R.~Paffenroth, B.~Sandstede, X.~Wang, C.~Zhang, Auto-07p:
	Continuation and bifurcation software for ordinary differential equations.
	
\end{thebibliography}

\end{document}